\documentclass[%
 reprint,
 amsmath,amssymb,
 aps,
 twocolumn, prd, superscriptaddress, showpacs, floatfix, nofootinbib 
]{revtex4-2}

\usepackage{graphicx}
\usepackage{dcolumn}
\usepackage{bm}
\usepackage{color}
\usepackage{hyperref}
\usepackage[mathlines]{lineno}
\usepackage{multirow}

\begin{document}

\preprint{APS/123-QED}

\title{Measurement of the cosmogenic neutron yield in Super-Kamiokande with gadolinium loaded water}



\newcommand{\AFFicrr}{\affiliation{Kamioka Observatory, Institute for Cosmic Ray Research, University of Tokyo, Kamioka, Gifu 506-1205, Japan}}
\newcommand{\AFFkashiwa}{\affiliation{Research Center for Cosmic Neutrinos, Institute for Cosmic Ray Research, University of Tokyo, Kashiwa, Chiba 277-8582, Japan}}
\newcommand{\AFFicrronly}{\affiliation{Institute for Cosmic Ray Research, University of Tokyo, Kashiwa, Chiba 277-8582, Japan}}
\newcommand{\AFFipmu}{\affiliation{Kavli Institute for the Physics and
Mathematics of the Universe (WPI), The University of Tokyo Institutes for Advanced Study,
University of Tokyo, Kashiwa, Chiba 277-8583, Japan }}
\newcommand{\AFFmad}{\affiliation{Department of Theoretical Physics, University Autonoma Madrid, 28049 Madrid, Spain}}
\newcommand{\AFFubc}{\affiliation{Department of Physics and Astronomy, University of British Columbia, Vancouver, BC, V6T1Z4, Canada}}
\newcommand{\AFFbu}{\affiliation{Department of Physics, Boston University, Boston, MA 02215, USA}}
\newcommand{\AFFuci}{\affiliation{Department of Physics and Astronomy, University of California, Irvine, Irvine, CA 92697-4575, USA }}
\newcommand{\AFFcsu}{\affiliation{Department of Physics, California State University, Dominguez Hills, Carson, CA 90747, USA}}
\newcommand{\AFFcnm}{\affiliation{Institute for Universe and Elementary Particles, Chonnam National University, Gwangju 61186, Korea}}
\newcommand{\AFFduke}{\affiliation{Department of Physics, Duke University, Durham NC 27708, USA}}
\newcommand{\AFFfukuoka}{\affiliation{Junior College, Fukuoka Institute of Technology, Fukuoka, Fukuoka 811-0295, Japan}}
\newcommand{\AFFgifu}{\affiliation{Department of Physics, Gifu University, Gifu, Gifu 501-1193, Japan}}
\newcommand{\AFFgist}{\affiliation{GIST College, Gwangju Institute of Science and Technology, Gwangju 500-712, Korea}}
\newcommand{\AFFuh}{\affiliation{Department of Physics and Astronomy, University of Hawaii, Honolulu, HI 96822, USA}}
\newcommand{\AFFicl}{\affiliation{Department of Physics, Imperial College London , London, SW7 2AZ, United Kingdom }}
\newcommand{\AFFkek}{\affiliation{High Energy Accelerator Research Organization (KEK), Tsukuba, Ibaraki 305-0801, Japan }}
\newcommand{\AFFkobe}{\affiliation{Department of Physics, Kobe University, Kobe, Hyogo 657-8501, Japan}}
\newcommand{\AFFkyoto}{\affiliation{Department of Physics, Kyoto University, Kyoto, Kyoto 606-8502, Japan}}
\newcommand{\AFFliv}{\affiliation{Department of Physics, University of Liverpool, Liverpool, L69 7ZE, United Kingdom}}
\newcommand{\AFFmiyagi}{\affiliation{Department of Physics, Miyagi University of Education, Sendai, Miyagi 980-0845, Japan}}
\newcommand{\AFFnagoya}{\affiliation{Institute for Space-Earth Environmental Research, Nagoya University, Nagoya, Aichi 464-8602, Japan}}
\newcommand{\AFFkmi}{\affiliation{Kobayashi-Maskawa Institute for the Origin of Particles and the Universe, Nagoya University, Nagoya, Aichi 464-8602, Japan}}
\newcommand{\AFFpol}{\affiliation{National Centre For Nuclear Research, 02-093 Warsaw, Poland}}
\newcommand{\AFFsuny}{\affiliation{Department of Physics and Astronomy, State University of New York at Stony Brook, NY 11794-3800, USA}}
\newcommand{\AFFokayama}{\affiliation{Department of Physics, Okayama University, Okayama, Okayama 700-8530, Japan }}
\newcommand{\AFFosaka}{\affiliation{Department of Physics, Osaka University, Toyonaka, Osaka 560-0043, Japan}}
\newcommand{\AFFox}{\affiliation{Department of Physics, Oxford University, Oxford, OX1 3PU, United Kingdom}}
\newcommand{\AFFqmul}{\affiliation{School of Physics and Astronomy, Queen Mary University of London, London, E1 4NS, United Kingdom}}
\newcommand{\AFFregina}{\affiliation{Department of Physics, University of Regina, 3737 Wascana Parkway, Regina, SK, S4SOA2, Canada}}
\newcommand{\AFFseoul}{\affiliation{Department of Physics, Seoul National University, Seoul 151-742, Korea}}
\newcommand{\AFFsheff}{\affiliation{Department of Physics and Astronomy, University of Sheffield, S3 7RH, Sheffield, United Kingdom}}
\newcommand{\AFFshizuokasc}{\affiliation{Department of Informatics in
Social Welfare, Shizuoka University of Welfare, Yaizu, Shizuoka, 425-8611, Japan}}
\newcommand{\AFFstfc}{\affiliation{STFC, Rutherford Appleton Laboratory, Harwell Oxford, and Daresbury Laboratory, Warrington, OX11 0QX, United Kingdom}}
\newcommand{\AFFskk}{\affiliation{Department of Physics, Sungkyunkwan University, Suwon 440-746, Korea}}
\newcommand{\AFFtokyo}{\affiliation{The University of Tokyo, Bunkyo, Tokyo 113-0033, Japan }}
\newcommand{\AFFtodai}{\affiliation{Department of Physics, University of Tokyo, Bunkyo, Tokyo 113-0033, Japan }}
\newcommand{\AFFtit}{\affiliation{Department of Physics,Tokyo Institute of Technology, Meguro, Tokyo 152-8551, Japan }}
\newcommand{\AFFtus}{\affiliation{Department of Physics, Faculty of Science and Technology, Tokyo University of Science, Noda, Chiba 278-8510, Japan }}
\newcommand{\AFFtoronto}{\affiliation{Department of Physics, University of Toronto, ON, M5S 1A7, Canada }}
\newcommand{\AFFtriumf}{\affiliation{TRIUMF, 4004 Wesbrook Mall, Vancouver, BC, V6T2A3, Canada }}
\newcommand{\AFFtokai}{\affiliation{Department of Physics, Tokai University, Hiratsuka, Kanagawa 259-1292, Japan}}
\newcommand{\AFFtsinghua}{\affiliation{Department of Engineering Physics, Tsinghua University, Beijing, 100084, China}}
\newcommand{\AFFynu}{\affiliation{Department of Physics, Yokohama National University, Yokohama, Kanagawa, 240-8501, Japan}}
\newcommand{\AFFllr}{\affiliation{Ecole Polytechnique, IN2P3-CNRS, Laboratoire Leprince-Ringuet, F-91120 Palaiseau, France }}
\newcommand{\AFFbari}{\affiliation{ Dipartimento Interuniversitario di Fisica, INFN Sezione di Bari and Universit\`a e Politecnico di Bari, I-70125, Bari, Italy}}
\newcommand{\AFFnapoli}{\affiliation{Dipartimento di Fisica, INFN Sezione di Napoli and Universit\`a di Napoli, I-80126, Napoli, Italy}}
\newcommand{\AFFroma}{\affiliation{INFN Sezione di Roma and Universit\`a di Roma ``La Sapienza'', I-00185, Roma, Italy}}
\newcommand{\AFFpadova}{\affiliation{Dipartimento di Fisica, INFN Sezione di Padova and Universit\`a di Padova, I-35131, Padova, Italy}}
\newcommand{\AFFkeio}{\affiliation{Department of Physics, Keio University, Yokohama, Kanagawa, 223-8522, Japan}}
\newcommand{\AFFwinnipeg}{\affiliation{Department of Physics, University of Winnipeg, MB R3J 3L8, Canada }}
\newcommand{\AFFkcl}{\affiliation{Department of Physics, King's College London, London, WC2R 2LS, UK }}
\newcommand{\AFFwarwick}{\affiliation{Department of Physics, University of Warwick, Coventry, CV4 7AL, UK }}
\newcommand{\AFFral}{\affiliation{Rutherford Appleton Laboratory, Harwell, Oxford, OX11 0QX, UK }}
\newcommand{\AFFwu}{\affiliation{Faculty of Physics, University of Warsaw, Warsaw, 02-093, Poland }}
\newcommand{\AFFbcit}{\affiliation{Department of Physics, British Columbia Institute of Technology, Burnaby, BC, V5G 3H2, Canada }}
\newcommand{\AFFtohoku}{\affiliation{Department of Physics, Faculty of Science, Tohoku University, Sendai, Miyagi, 980-8578, Japan }}
\newcommand{\AFFicise}{\affiliation{Institute For Interdisciplinary Research in Science and Education, ICISE, Quy Nhon, 55121, Vietnam }}
\newcommand{\AFFilance}{\affiliation{ILANCE, CNRS - University of Tokyo International Research Laboratory, Kashiwa, Chiba 277-8582, Japan}}
\newcommand{\AFFibs}{\affiliation{Institute for Basic Science (IBS), Daejeon, 34126, Korea}}

\AFFicrr
\AFFkashiwa
\AFFicrronly
\AFFmad
\AFFbu
\AFFbcit
\AFFuci
\AFFcsu
\AFFcnm
\AFFduke
\AFFllr
\AFFfukuoka
\AFFgifu
\AFFgist
\AFFuh
\AFFibs
\AFFicise
\AFFicl
\AFFbari
\AFFnapoli
\AFFpadova
\AFFroma
\AFFilance
\AFFkeio
\AFFkek
\AFFkcl
\AFFkobe
\AFFkyoto
\AFFliv
\AFFmiyagi
\AFFnagoya
\AFFkmi
\AFFpol
\AFFsuny
\AFFokayama
\AFFox
\AFFral
\AFFseoul
\AFFsheff
\AFFshizuokasc
\AFFstfc
\AFFskk
\AFFtohoku
\AFFtokai
\AFFtokyo
\AFFtodai
\AFFipmu
\AFFtit
\AFFtus
\AFFtoronto
\AFFtriumf
\AFFtsinghua
\AFFwu
\AFFwarwick
\AFFwinnipeg
\AFFynu

\author{M.~Shinoki}
\AFFtus
\author{K.~Abe}
\AFFicrr
\AFFipmu
\author{Y.~Hayato}
\AFFicrr
\AFFipmu
\author{K.~Hiraide}
\AFFicrr
\AFFipmu
\author{K.~Hosokawa}
\AFFicrr
\author{K.~Ieki}
\author{M.~Ikeda}
\AFFicrr
\AFFipmu
\author{J.~Kameda}
\AFFicrr
\AFFipmu
\author{Y.~Kanemura}
\author{R.~Kaneshima}
\author{Y.~Kashiwagi}
\AFFicrr
% ...


\collaboration{The Super-Kamiokande Collaboration}
\noaffiliation


\begin{abstract}
Cosmic-ray muons that enter the Super-Kamiokande detector cause hadronic showers due to spallation in water, producing neutrons and radioactive isotopes. These are a major background source for studies of MeV-scale neutrinos and searches for rare events. In 2020, gadolinium was introduced into the ultra-pure water in the Super-Kamiokande detector to improve the detection efficiency of neutrons. In this study, the cosmogenic neutron yield was measured using data acquired during the period after the gadolinium loading. The yield was found to be $(2.76 \pm 0.02\,\mathrm{(stat.) \pm 0.19\,\mathrm{(syst.)}}) \times 10^{-4}\,\mu^{-1} \mathrm{g^{-1} cm^{2}}$ at an average muon energy 259\,GeV at the Super-Kamiokande detector.
\end{abstract}


\maketitle


\section{\label{sec:intro}Introduction}
High-energy muons are produced in the atmosphere from the interactions of cosmic rays and penetrate deep underground.
Muons, or electromagnetic showers caused by muons, interact with nuclei to produce secondary particles consisting of nucleons and mesons.
These secondary particles are produced when a muon interacts with a nucleus via a virtual photon, causing a photodisintegration, or when a nucleus absorbs a $\gamma$ ray from an electromagnetic shower caused by a muon \cite{malgin, Luu2006NeutronPB, wang_etal, calc_li_beacom}. Furthermore, secondary particles interact with nuclei and subsequently produce neutrons and unstable radioactive isotopes by spallation processes.
In the energy range of muons reaching Super-Kamiokande (SK), neutrons and isotopes are dominantly produced by the interactions of $\pi^-$ and nucleons \cite{wang_etal, calc_li_beacom}.
Neutrons produced by spallation are captured after thermalization, and $\gamma$ rays with energies of several MeV are emitted, while radioactive isotopes decay with MeV-scale $\beta$ or $\beta \gamma$.
It is important to understand these spallation processes as they constitute one of the major backgrounds for solar neutrinos and the diffuse supernova neutrino background \cite{calc_li_beacom2}.
Due to the nature of these complicated spallation processes, many underground experiments rely on their measurements to estimate the background contamination in the searches. It is important for future projects to understand their production mechanism to improve the precision of the background estimation. In this analysis, muon-induced neutrons are measured in SK for the first time with the gadolinium loaded water.

Several experiments have measured cosmogenic neutron production yields at various depths, mostly with liquid scintillator-based detectors \cite{hertenberger,boehm,aberdeen,dayabay,kamland,lvd,borexino}. Among these, the KamLAND detector is located at about the same depth as the SK detector \cite{kamland},
but the liquid scintillator is pseudocumene based and therefore contains mostly carbon and hydrogen.
Therefore, a comparison of the neutron yields in SK and KamLAND provides unique information to investigate the dependence on the atomic number for muons with similar energy spectra.

This paper describes the measurement of neutrons produced by the spallation of cosmic-ray muons in SK. The overview of the SK detector and trigger system is explained in Sec.~\ref{sec:sk_experiment}. The detector simulation is described in Sec.~\ref{sec:simulation}. The analysis methods are explained in  Sec.~\ref{sec:analysis} where muon selection, neutron detection, and the systematic uncertainty are discussed. The results of the neutron yield measurement and comparisons with other experiments are presented in Sec.~\ref{sec:results}. Finally, we present our conclusion in Sec.~\ref{sec:conclusion}.

\section{\label{sec:sk_experiment}Super-Kamiokande}

The SK detector is a large water Cherenkov detector installed 1000\,m underground (2700\,meter water equivalent) in Kamioka, Japan \cite{Super-Kamiokande:2002weg}.
The detector is a cylindrical tank with a diameter of 39.3\,m and a height of 41.4\,m. The tank is filled with about 50\,kton of gadolinium (Gd) doped ultra-pure water \cite{sk_gd_loading}.
The mass concentration of Gd is 0.011 wt\%.
The SK detector is divided into two concentric volumes: an inner detector (ID) and an outer detector (OD).
The ID is a cylindrical volume with a diameter of 33.8 m and a height of 36.2 m. It is surrounded by 11129 inward-facing 20-in photomultiplier tubes (PMTs).
The OD surrounds the ID with a thickness of 2.05\,m on the top and bottom and 2.2\,m on the sides.
There are 1885 outward-facing 8-in PMTs attached to the walls of the OD.
The main purpose of the OD is to identify cosmic-ray muons and to attenuate $\gamma$ rays and neutrons produced in the surrounding rock. When the number of OD PMT hits within 200\,ns exceeds 22, an OD trigger is issued.

Events are triggered by the total number of coincidence ID PMT hits within 200\,ns.
If an event with 58 or more ID PMT hits, 
which corresponds to an electron equivalent energy deposit of $\sim$7.5\,MeV near the center of the detector, is triggered, all PMT hits for 535\,$\mu$s after the trigger are recorded.
Therefore, once a cosmic-ray muon is triggered, signals from neutron capture due to the spallation can be searched for up to 535\,$\mu$s by offline analysis with the lower threshold than that of normal trigger. These delayed signals at the MeV scale after the muon are called ``low-energy events'' in this analysis.

SK operated with ultra-pure water from April 1996 to July 2020, a period consisting of five phases.
In the fourth phase (SK-IV), new front-end electronics and a new trigger system were introduced to allow neutron tagging and increase data throughput \cite{sk_qbee}. The SK-IV phase had the longest operational period, 2970\,days, which continued until the start of the refurbishment work in May 2018.
In ultra-pure water, a neutron is captured by a hydrogen nucleus after about 200\,$\mu$s on average and this process emits a $\gamma$ ray with an energy of 2.2\,MeV. Although this energy is lower than the trigger threshold in SK, neutrons have been tagged by an analysis using machine learning methods \cite{sk_li9, sk_1st_ntag, sk4_srn, ntag_sk_2022}.
The neutron detection efficiency was 20--25\% in SK-IV.
The dissolution of gadolinium sulfate octahydrate $\mathrm{Gd_2(SO_4)_3\cdot8H_2O}$ in the ultra-pure water started in July 2020 after a short period with ultra-pure water operation (SK-V) following the tank refurbishment work completed in 2019. 12.9\,tons of $\mathrm{Gd_2(SO_4)_3\cdot8H_2O}$ was loaded in 50\,kton of ultra-pure water, corresponding to a mass concentration of 0.011\,wt\% Gd, in August 2020 and SK-VI running period has started \cite{sk_gd_loading}.
Gadolinium has a large neutron capture cross section and $\sim$50\% neutrons are captured on Gd after $\sim$116\,$\mu$s on average with this Gd concentration.
Several $\gamma$ rays totaling about 8\,MeV are emitted after neutron capture by Gd. This can be clearly distinguished from the background consisting of environmental radiation and dark noise of PMTs. Therefore, the neutron-tagging efficiency is significantly improved by loading Gd in water.
In this paper, 283.2\,days of data taken during SK-VI from September 2020 to September 2021 is analyzed after the Gd concentration became uniform throughout the detector tank.
Details of the Gd loading and the detector status were described in Ref.~\cite{sk_gd_loading}.

\section{\label{sec:simulation}Detector simulation}
Monte Carlo (MC) simulation is used in this study to calculate the signal efficiencies for event selections and evaluate the systematic uncertainties when data-driven estimation is difficult. This simulation is based on GEANT3 \cite{GEANT3}. The simulations take into account the detector geometry and water quality, particle propagation in water, Cherenkov light emission, light absorption and scattering, and PMT and electronics response \cite{sk_calib}. In addition, another simulation based on GEANT4 \cite{GEANT4:2002zbu} is used to model the process of neutron propagation in water with energies below 20\,MeV, which includes neutron capture reactions by nuclei, and $\gamma$-ray emissions. For the multiplicity and energy spectrum of $\gamma$ rays in the thermal neutron capture reaction of ${}^{155}$Gd and ${}^{157}$Gd, a model reflecting the results measured at the ANNRI neutron beam line at J-PARC/MLF \cite{annri_gd} is incorporated.
In this study, neutron capture events are simulated by generating single neutrons uniformly throughout the ID.
Background samples containing the dark noise and the radio activities in the detector are collected from the data using periodic triggers and added to the simulation.
The trigger search is performed on the simulation results by the same algorithm as the data, and the reconstruction is applied to each triggered event.

\section{\label{sec:analysis}Analysis}
Muon events and low-energy events described in the following sections are selected as the candidates of neutrons induced by the cosmic-ray muon spallation and captured by Gd.

\subsection{\label{sec:cosmic_ray_muon}Cosmic-ray muons}
Cosmic-ray muons are recorded by both ID and OD triggers.
In this analysis, muon candidates are selected by requiring the total number of observed photoelectrons of the ID PMTs to be greater than 1000, which corresponds to $\sim$140\,MeV, and the tracks are reconstructed with the muon fitter.
The details of the algorithm for the muon fitter were described in Ref.~\cite{muboy_conner, muboy_desai}.
The cosmic-ray muons are classified into four types; single through-going (88.8\% of all muons), stopping (3.9\%), multiple (7.3\%), and corner-clipping (0.003\%). 
A single through-going muon has a single track and penetrates the ID. If a single muon loses energy and stops inside the ID, it is classified as the stopping type. When muon bundles pass through the ID, they are categorized as the multiple type.
A single muon that grazes the edge of the ID is classified as the corner-clipping type.
The total number of muons used in this analysis, $N_\mu$, is counted as $N_\mu = 4.77 \times 10^7$ for 283.2\,days of exposure, corresponding to 1.95\,muons/s.
\begin{figure}
    \centering
    \includegraphics[width=8.5cm]{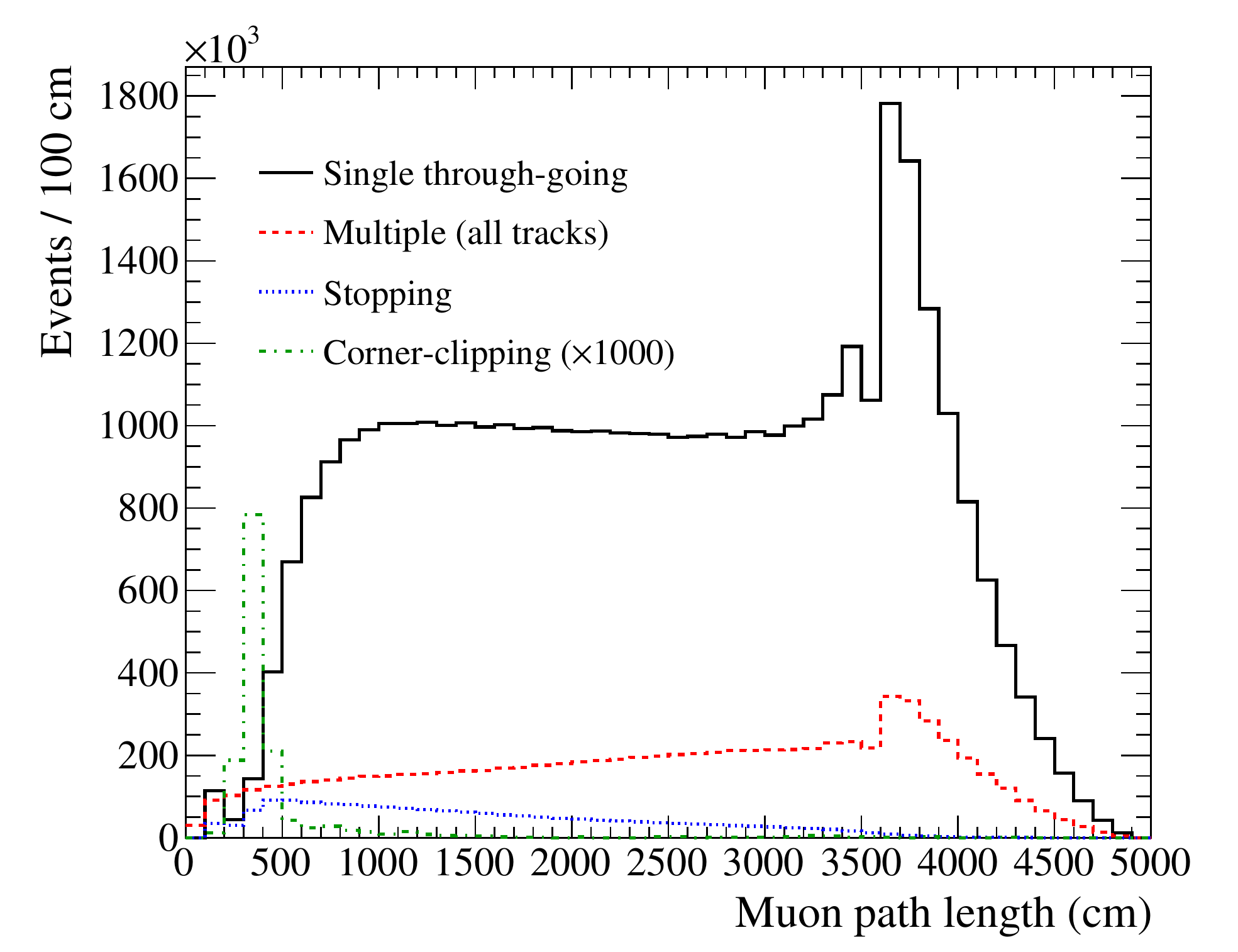}
    \caption{\label{fig:muon_path_length} Distributions of path length $L_\mu$ for single through-going (solid), multiple (dashed), stopping (dotted), and corner-clipping (dot-dashed) muons. The distribution of corner-clipping muons is scaled by a factor of 1000.}
\end{figure}

Figure~\ref{fig:muon_path_length} shows the distributions of the track lengths in the detector.
The peak of the track length distribution at around 3700\,cm corresponds to muons that penetrate both the top and bottom end caps traveling straight down through the detector.
The average path length $L_\mu$ was obtained to be $L_\mu = 2427$\,cm.

\begin{figure}
    \centering
    \includegraphics[width=8.5cm]{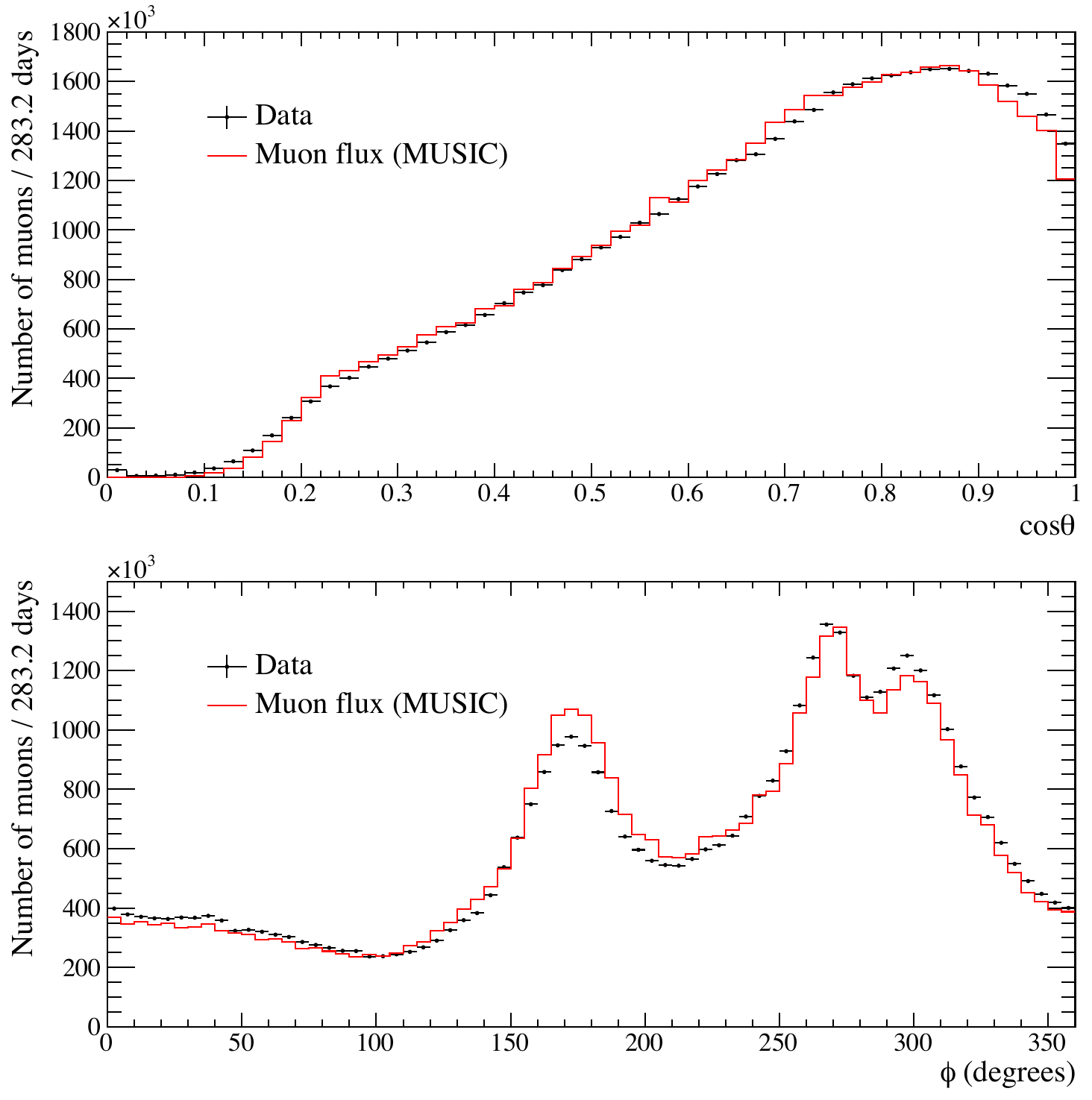}
    \caption{\label{fig:muon_flux} Zenith ($\theta$) and azimuthal ($\phi$) dependence of  the muon rate at the SK site. Reconstructed directions of the muons from the data (black points) are overlaid with the muon flux at the SK site calculated by the MUSIC code (red lines). Muon flux distributions are normalized to the data. $\phi = 0$ corresponds to the direction from east to west.}
\end{figure}

The number of neutrons produced in muon spallation depends on the energy of the muon.
The energy of cosmic-ray muons at the SK site is estimated using simulation.
The muon flux at sea level is modeled by modifying Gaisser's parametrization \cite{gaisser} according to Ref.~\cite{tang}.
The MUSIC code \cite{music} is used to simulate muon propagation in the rock.
The simulations account for the topography of Mount Ikenoyama surrounding the SK area \cite{diggitalmap, kamland} and the rock models of standard \cite{groom, barrett} and Ikenoyama \cite{tang}.
The density of the rock is assumed to be 2.65--2.75\,$\mathrm{g/cm^3}$. The average muon energy incident on the SK detector $\overline{E}_\mu$ is estimated to be $\overline{E}_\mu = 259 \pm 9$\,GeV with the calculation method from Ref.~\cite{tang}, where the uncertainty was estimated by varying the rock model and density. Figure~\ref{fig:muon_flux} shows the cosine of the zenith angle and azimuthal angle of muons at the SK site. The muon flux from the MUSIC code is overlaid with the reconstructed directions of the data.

\subsection{\label{subsec:event_selection}Neutron capture event selection}
To select neutron captures on Gd, several cuts are applied to the low-energy events following the muon.
\subsubsection{\label{subsec:basic_reduction}Basic reduction}
The time difference between the muon and the following events are defined as $\Delta t$.
Neutron capture candidates are searched within the time window $40\,\mathrm{\mu s} \leq \Delta t \leq 530\,\mathrm{\mu s}$. The time $\Delta t < 40\,\mathrm{\mu s}$ is not used to avoid the contamination of the decay electron from a muon and PMT afterpulses.

In order to exclude background due to radioactive decay near the ID wall \cite{sk_nakano_Rn} and neutrons entering from the surrounding rock, a fiducial volume cut is applied based on the reconstructed vertex position. The fiducial volume of this analysis is defined with the boundary 4\,m away from the ID wall, and the ratio of the fiducial volume to the total volume of the ID is accounted for as the signal efficiency of the volume cut.
The fiducial volume is smaller than the other analysis in SK \cite{sk4_srn} in order to suppress the systematic uncertainty due to leak-in/-out of neutrons, as described in Sec.\ref{subsection:syst_uncertainty}, while the statistical uncertainty is still smaller than the systematic uncertainty.

\subsubsection{\label{subsec:event_quality_cut}Event quality}
Unlike signal events, which have a peaked timing distribution and a ring pattern, the time and location of PMT hits for background events are randomly distributed within the detector.
Therefore, event reconstruction often does not work well. Such background can be reduced by evaluating the goodness of event reconstruction \cite{smy_proc}.
The timing information of the hit PMTs is used to reconstruct the event vertex. For the $i$th hit PMT, a residual time $\Delta \tau_i$ is defined as
\begin{equation}
    \Delta \tau_i = t_i - t_\mathrm{tof} - t_0,
\end{equation}
where $t_i$ is the time when the signal was detected, $t_\mathrm{tof}$ is the time of flight of the photon to reach the hit PMT from the event vertex, and $t_0$ is the time of the $\gamma$ ray emission from the neutron capture.
The event vertex is determined by minimizing the width of the $t_i - t_\mathrm{tof}$ distribution.
The parameter $g_t$, which represents the degree of certainty of the vertex reconstruction, is defined as
\begin{equation}
    g_t = \frac{ \sum_i \exp \left[ -\frac{1}{2} \left( \frac{\Delta \tau_i}{\omega} \right)^2 \right] \exp \left[ -\frac{1}{2} \left( \frac{\Delta \tau_i}{\sigma} \right)^2 \right] }{ \sum_i \exp \left[ -\frac{1}{2} \left( \frac{\Delta \tau_i}{\omega} \right)^2 \right] },
\end{equation}
where $\omega$ gives the weight to suppress the dark noise and $\sigma$ is the time resolution of the PMT for a single photoelectron signal, which are set to 60 and 5\,ns, respectively.

The event direction is reconstructed using the maximum-likelihood method, which finds the Cherenkov ring that best matches the positions of the hit PMTs. 
The goodness-of-direction reconstruction, $g_p$, is the Kolmogorov–Smirnov statistic from a comparison between the observed hit PMTs and the expectation assuming $\phi$ symmetry around the reconstructed direction \cite{koshio}.

\begin{figure}
    \centering
    \includegraphics[width=8.5cm]{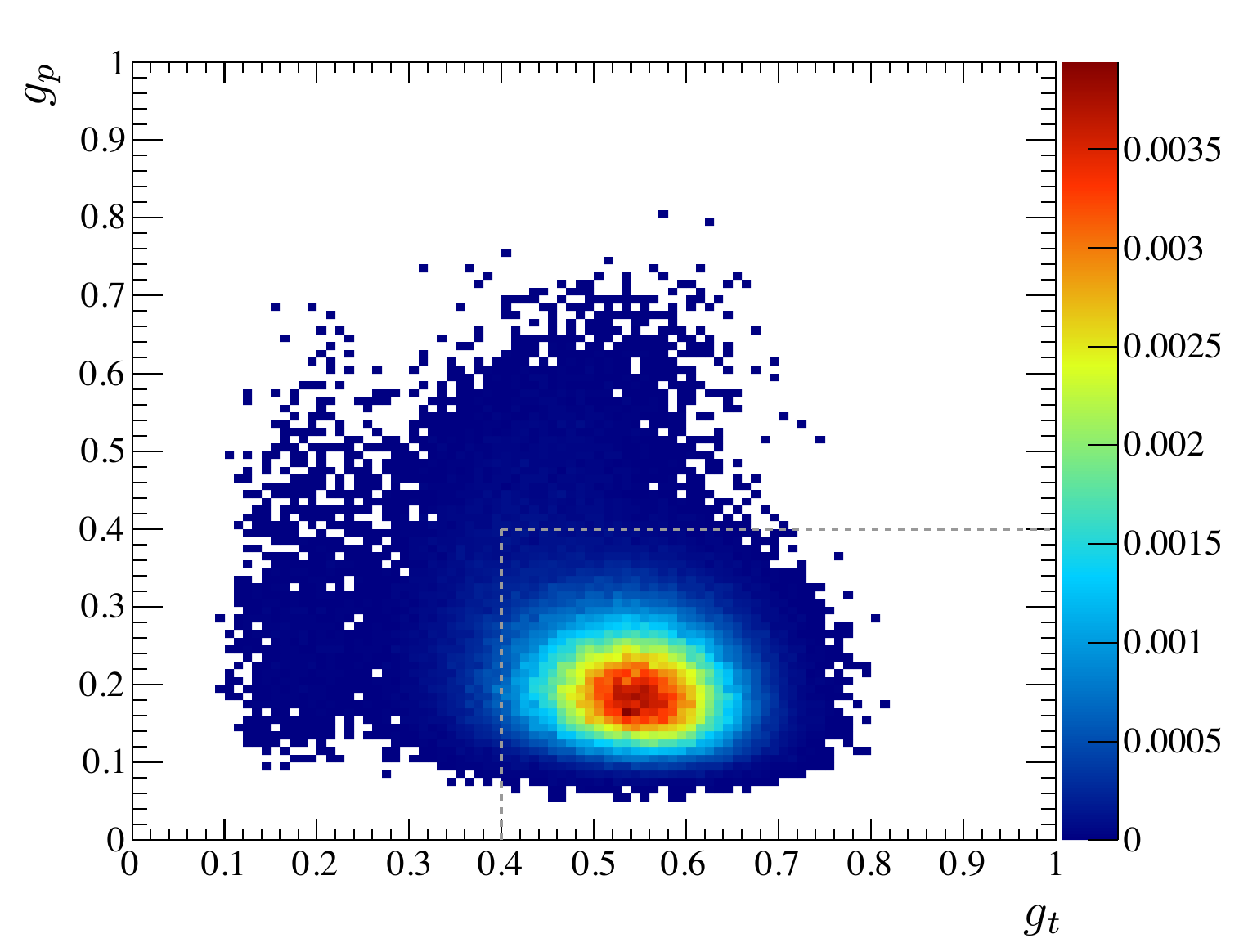}
    \caption{\label{fig:bsgood_dirks_mc} Distribution of $g_t$ and $g_p$ for neutron capture on Gd in the MC. The color scale is the number of events normalized by area. The gray dashed lines are the cut thresholds.}
\end{figure}

$g_t$ ($g_p$) is a variable that takes values close to one (zero) for higher degrees of confidence in the reconstruction.
The distribution of $g_t$ and $g_p$ for neutron capture on Gd in the MC is shown in Fig.~\ref{fig:bsgood_dirks_mc}.
In this analysis, events with $g_t > 0.4$ and $g_p < 0.4$ are retained as signal candidates. The signal efficiency for the event quality cut is evaluated with the MC as $(92.62 \pm 0.29)$\% with the MC statistical uncertainty.

\subsubsection{\label{subsec:n50_cut}Number of hit PMTs}
\begin{figure}
    \centering
    \includegraphics[width=8.5cm]{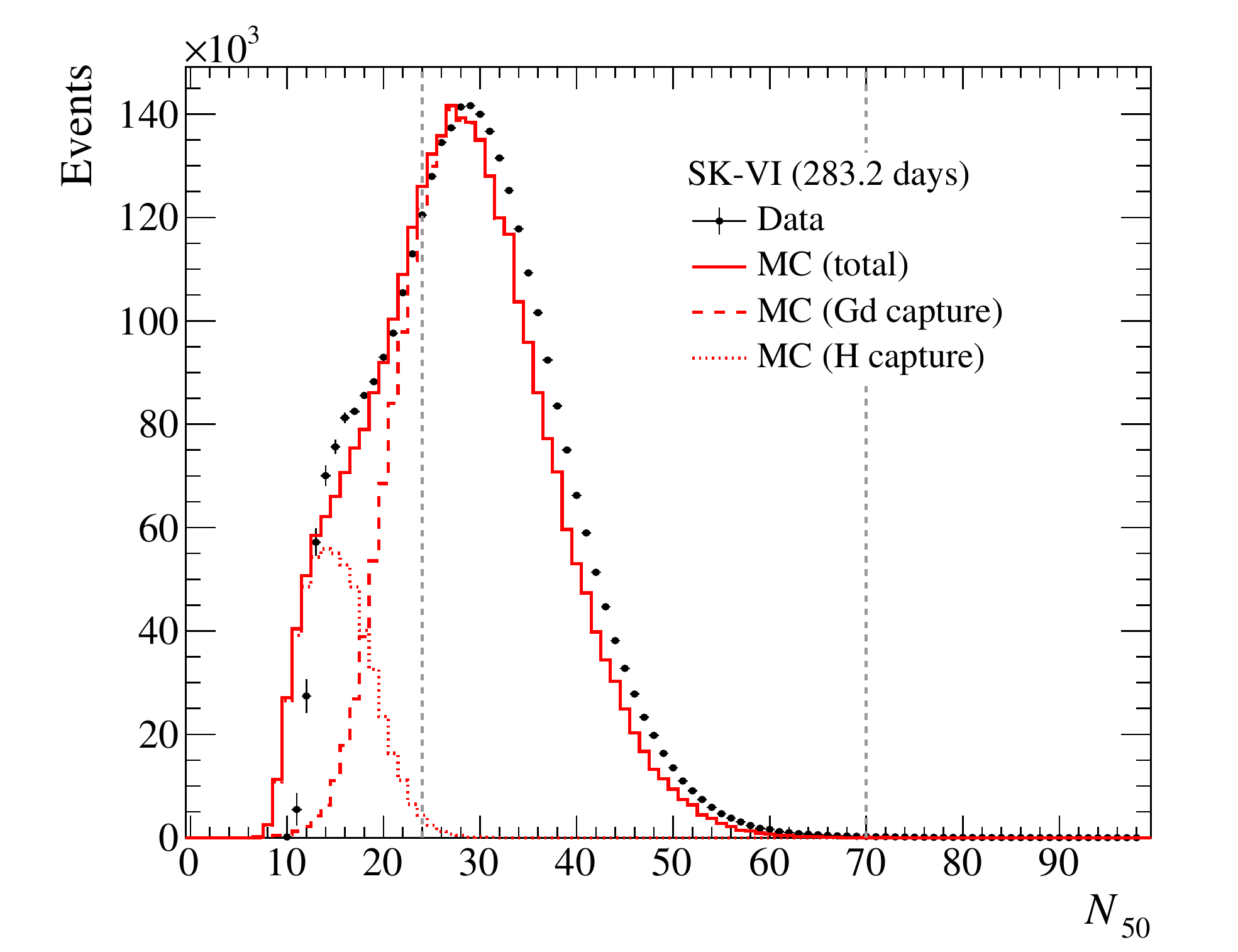}
    \caption{\label{fig:n50} Distribution of $N_{50}$ for data (black points, background subtracted) and MC (red lines). Total (solid line), Gd capture (dashed line), and hydrogen capture (dotted line) are plotted. All cuts except to the $N_{50}$ cut are applied for the data. The remaining background is evaluated from off-time ($430\,\mathrm{\mu s} \leq \Delta t \leq 530\,\mathrm{\mu s}$) and subtracted from on-time ($40\,\mathrm{\mu s} \leq \Delta t \leq 240\,\mathrm{\mu s}$). Only the event quality cut is applied for MC and the entries of the total MC distribution are normalized to the data by the height of the peak. The gray dashed lines are $N_{50}$ cut thresholds.}
\end{figure}

Energy reconstruction is based on the number of hit PMTs as most of them are single photoelectron signals. Since the background due to the radioactivities exists dominantly at low energy below the signal from neutron capture on Gd, event selection is applied by setting a threshold for the number of hit PMTs.
The time of flight of photons from the reconstructed vertex to the hit PMTs is subtracted from the PMT hit time. The number of hit PMTs in a 50-ns time window is then defined as $N_{50}$.
The $N_{50}$ distributions for both data and the MC are shown in Fig.~\ref{fig:n50}. Background events due to accidental coincidence are evaluated from the off-time window and subtracted.
In order to suppress the contamination of the hydrogen capture events, the cut criterion is set to $24 \leq N_{50} \leq 70$.
The signal efficiency for $N_{50}$ cut is obtained as $(80.22 \pm 0.27)$\% using the distribution of neutron captures in the MC. The uncertainty is due to the MC statistics. The explanation of systematic uncertainty is described in Sec.~\ref{subsection:syst_uncertainty}.

\subsubsection{\label{subsec:lt_cut}Distance from muon track}
\begin{figure}
    \centering
    \includegraphics[width=5.5cm]{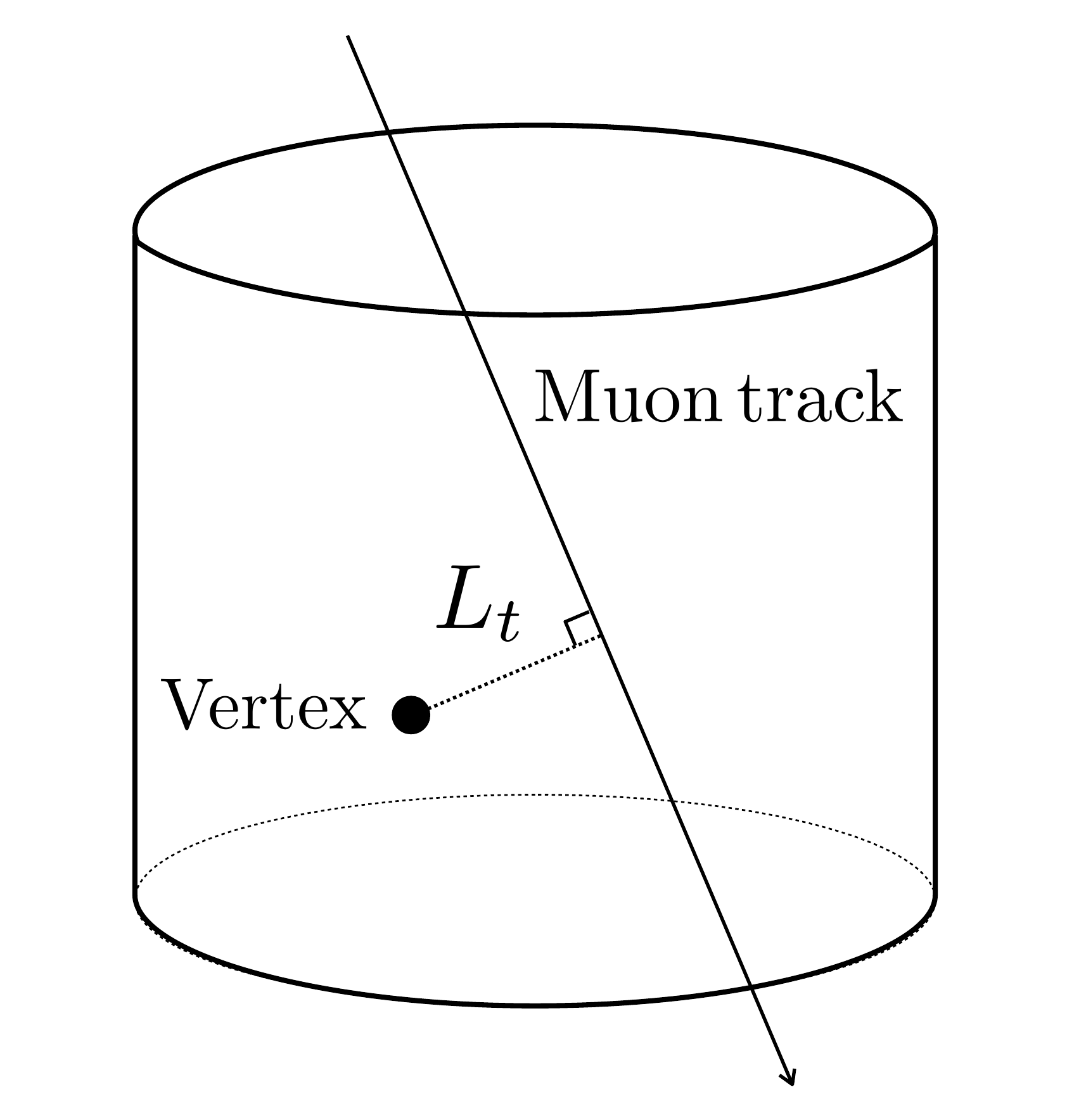}
    \caption{\label{fig:lt_def} Definition of transverse distance ($L_t$) between the muon track and the reconstructed vertex.}
\end{figure}

\begin{figure}
    \centering
    \includegraphics[width=8.5cm]{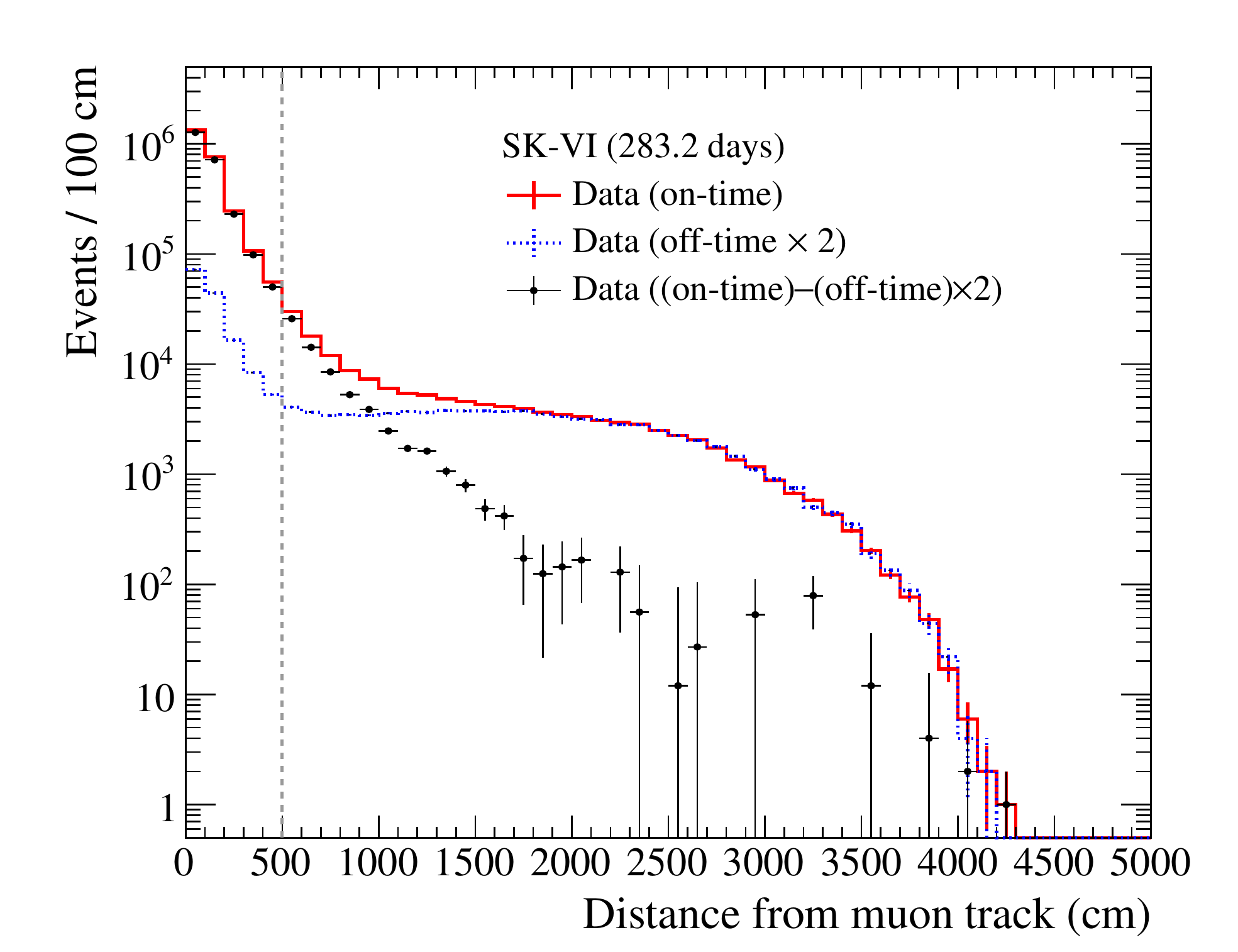}
    \caption{\label{fig:lt_data} Distribution of transverse distance ($L_t$) between the muon track and the reconstructed vertex for data (black points). The background is evaluated from off-time ($430\,\mathrm{\mu s} \leq \Delta t \leq 530\,\mathrm{\mu s}$, dotted line) and subtracted from on-time ($40\,\mathrm{\mu s} \leq \Delta t \leq 240\,\mathrm{\mu s}$, solid line). Since neutrons are captured on Gd after $\sim$100\,$\mathrm{\mu s}$ on average, events also occur in the off-time range and a peak is seen within 500\,cm. The dashed line is the $L_t$ cut threshold.}
\end{figure}

To select muon-induced neutrons, it is effective to use the transverse distance between the muon track and the reconstructed vertex. The definition of the transverse distance $L_t$ is shown in Fig.~\ref{fig:lt_def}. $L_t$ correlates with the distance that the secondary particles produced by the muon-induced hadronic shower travel through water. For multiple muons, $L_t$ is defined as the distance between the reconstructed vertex and the closest muon track. The $L_t$ distribution of the data is shown in Fig.~\ref{fig:lt_data} for events after all cuts except for the $L_t$ cut. In this analysis, the selection criterion is determined as $L_t < 500\,\mathrm{cm}$ and the signal efficiency is evaluated as $(97.25 \pm 0.10)$\% from the data after subtraction of off-time.

\subsection{\label{subsec:number_of_neutrons}Number of neutrons}
\begin{figure}
    \centering
    \includegraphics[width=8.5cm]{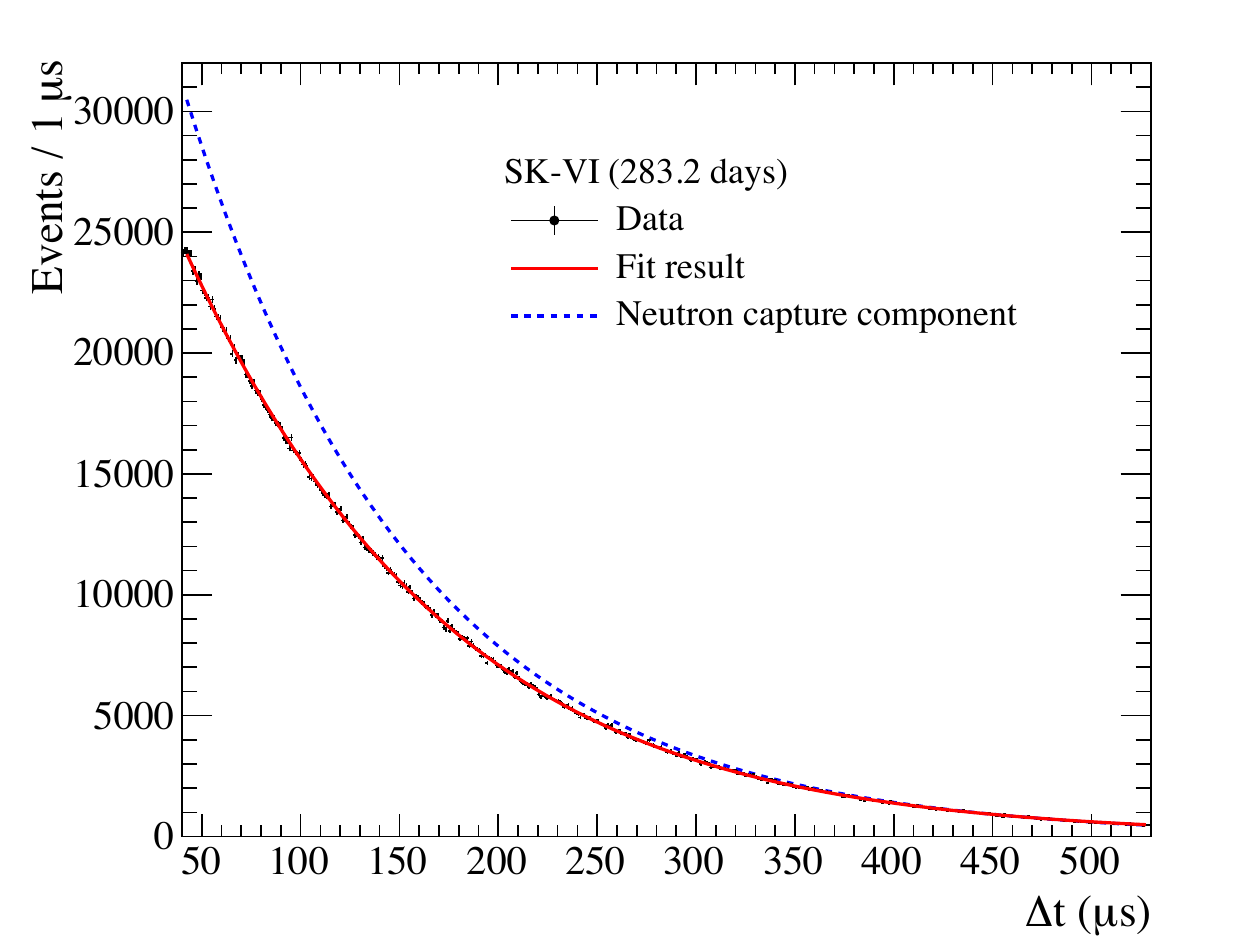}
    \caption{\label{fig:dt_fit} Distribution of the time difference between the muon and the following neutron capture candidates. The data are shown by filled circles.
    The solid red line is the fit result [Eq.~(\ref{eq:fitfunc_deltat})]. The dashed blue line is the neutron capture component.
    The reduced $\chi^2$ is found to be $\chi^2/\mathrm{d.o.f.} = 500.6/487$ as the fitting result, corresponding to a $p$-value of 32.5\%.
    }
\end{figure}

\begin{figure}
    \centering
    \includegraphics[width=8.5cm]{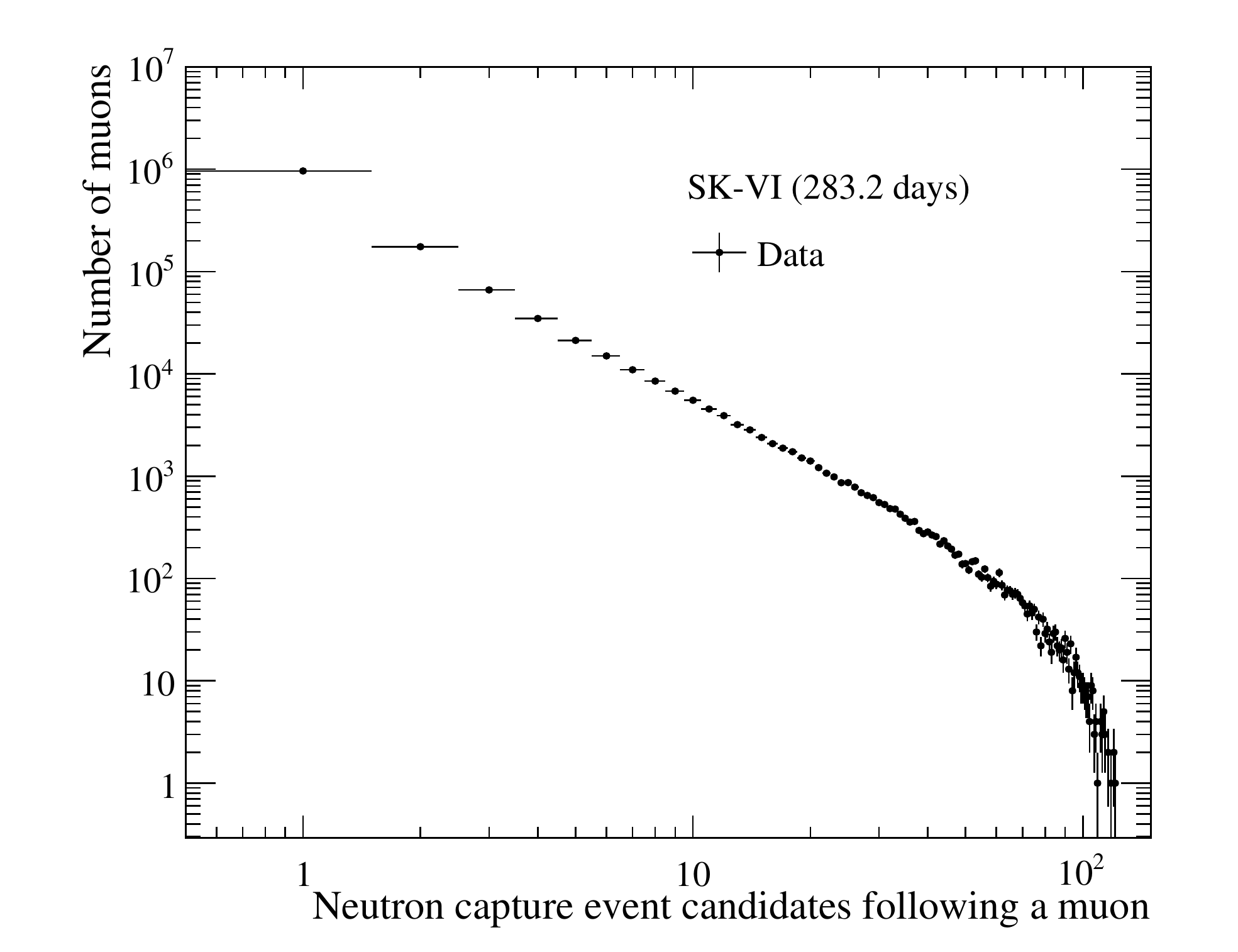}
    \caption{\label{fig:neutron_multiplicity} Distribution of the number of neutron candidates following a muon that remain after the event selections.}
\end{figure}

The total number of neutron capture signals is extracted by using the time difference between the muon and the following neutron candidates $\Delta t$. The $\Delta t$ distribution is shown in Fig.~\ref{fig:dt_fit}.
The capture time of neutrons in SK is measured using the americium beryllium (Am/Be) source \cite{sk_gd_loading}. It should be taken into account that a single neutron is emitted from the Am/Be source, while several neutrons are often emitted from muon spallation.
Figure \ref{fig:neutron_multiplicity} shows the number of neutron capture candidates observed following a muon that remain after the event selections.
Neutron capture occurs frequently over short time intervals in case several neutrons are produced in the detector by the muon spallation.
Neutron capture candidates are searched with sliding time windows and the PMT hit information for 1.3\,$\mu$s around the candidate is identified as a single event. If there are multiple neutron captures within 1.3\,$\mu$s, only one of them is selected as a signal candidate.
Since the effect of the dead time follows an exponential function, the $\Delta t$ distribution is fitted with the function
\begin{equation}
    f(\Delta t) = \left(A \exp \left( - \frac{\Delta t}{\tau_n} \right) + B \right) \left(1 - C \exp \left( - \frac{\Delta t}{\tau_\mathrm{d}} \right) \right),
    \label{eq:fitfunc_deltat}
\end{equation}
where $\tau_n$ is the neutron capture time constant obtained from measurements of Am/Be calibration, $\tau_n = 116.4 \pm 0.3\,\mathrm{\mu s}$. $B$ represents the background events and is fixed to be 27.80, which was determined using the data obtained with random triggers.
The first part of Eq.~(\ref{eq:fitfunc_deltat}) corresponds to the neutron capture component and the second part represents the inefficiency due to the dead time effect.
The uncertainty on the background estimation is accounted as the systematic uncertainty (Sec.~\ref{subsection:syst_uncertainty}). $A$ and $C$ are parameters determined by a fit. $A$ represents the normalization of neutron events. The second term absorbs the effect of dead time with the time constant $\tau_\mathrm{d}$ with $C$ as the normalization parameter.
The best-fit parameters are $A = (4.39 \pm 0.02) \times 10^4$,  $C = 0.256 \pm 0.003$, and $\tau_\mathrm{d} = 216.6 \pm 10.0\,\mathrm{\mu s}$.
The total number of detected neutrons from captures, $S_n$ (defined hereafter as signal neutrons), is obtained as $S_n = (3.57 \pm 0.02) \times 10^6$ using the integral
\begin{equation}
    S_n = \int_{40\,\mathrm{\mu s}}^{530\,\mathrm{\mu s}} A \exp \left( - \frac{\Delta t}{\tau_n} \right) d(\Delta t).
\end{equation}

In order to evaluate the signal efficiency for the time window, the function is obtained by the fit to the region $\Delta t < 40\,\mathrm{\mu s}$.
The time for the neutrons to thermalize in water is taken into account in the following function
\begin{equation}
    f_\mathrm{th}(\Delta t) = A \left( 1 - \exp \left( - \frac{\Delta t}{\tau_\mathrm{th}} \right) \right) \exp \left( - \frac{\Delta t}{\tau_n} \right),
\end{equation}
where $\tau_\mathrm{th}$ is the time constant for thermalization of neutrons in water, $\tau_\mathrm{th} = 4.3\,\mathrm{\mu s}$ \cite{sk_gd_loading}. From the ratio of integrals from 40 to 530\,$\mathrm{\mu s}$ to integrals from zero to infinity for the function $f_\mathrm{th}(\Delta t)$, the signal efficiency for the time window is estimated to be $(72.45 \pm 0.39)$\%.

The total number of neutrons produced by the muons can be obtained by correcting $S_n$ using the signal efficiency:
\begin{equation}
    N_n = \frac{S_n}{\epsilon},
\end{equation}
where $N_n$ is the total number of neutrons and $\epsilon$ is the signal efficiency.
The signal efficiencies for each event selection are summarized in Table~\ref{tab:summary_efficiency}.
\begin{table}
\caption{\label{tab:summary_efficiency}Summary of the signal efficiencies for each event selection. The errors are statistical only.}
\begin{ruledtabular}
\begin{tabular}{lc}
    Event selection & Efficiency (\%)\\
    \colrule
    Analysis volume & \multirow{2}{*}{45.39}\\
    (with respect to the ID of 32.5\,kton) & \\
    Event quality & $92.62 \pm 0.29$\\
    $N_{50}$ & $80.22 \pm 0.27$\\
    $L_t$ & $97.25 \pm 0.10$\\
    Time window & $72.45 \pm 0.39$\\
    Gd capture & 47\\
    \colrule
    Total & $11.17 \pm 0.08$
\end{tabular}
\end{ruledtabular}
\end{table}
The neutron capture fraction on Gd depends on the Gd concentration in water and the neutron capture time constant. The Gd capture fraction is estimated to be $47$\% from the correlation with the neutron capture time constant given by the GEANT4-based simulation, and the uncertainty is estimated as 1\%, which corresponds to a systematic uncertainty in $N_n$ of 2.2\%.
Signal efficiency including all effects was obtained as $(11.17 \pm 0.08)$\%.

\subsection{\label{subsection:syst_uncertainty}Systematic uncertainty}
The systematic uncertainties are summarized in Table~\ref{tab:summary_syst}.
The uncertainties on the number of muons and muon path length come from the performance of the muon fitter. For multiple muons, the number of simultaneously incident muons is counted up to ten and reconstructed assuming that they penetrate the ID in parallel.
To validate the accuracy of the muon fitter’s counting, the event displays are checked and the number of muon tracks is counted. The discrepancy from the muon fitter result is 2.0\% and this value is accounted for as the systematic uncertainty on $N_\mu$ due to the muon fitter accuracy.
The systematic uncertainty on the neutron yield due to the measurement of $L_\mu$ is estimated to be 1.3\% from the accuracy of the path length reconstruction, which is about 30\,cm.

The uncertainty for the Gd capture fraction was discussed in Sec.~\ref{subsec:number_of_neutrons}.
The systematic uncertainty on the Gd capture time is estimated as 1.2\% by considering the uncertainty on $\tau_n$.
The systematic uncertainty due to the model of neutron thermalization is estimated to be 2.2\% by using simulations assuming different thermal scattering processes.
The systematic uncertainty of the $L_t$ cut is evaluated by applying different $L_t$ cuts with the thresholds ranging from 400 to 600\,cm. The relative variation is below 0.6\%, which is assigned as the systematic uncertainty. This range of the variation of the threshold comes from the resolutions of the entrance position of the muon reconstruction and the vertex position of the low-energy events.

\begin{figure}
    \centering
    \includegraphics[width=8.5cm]{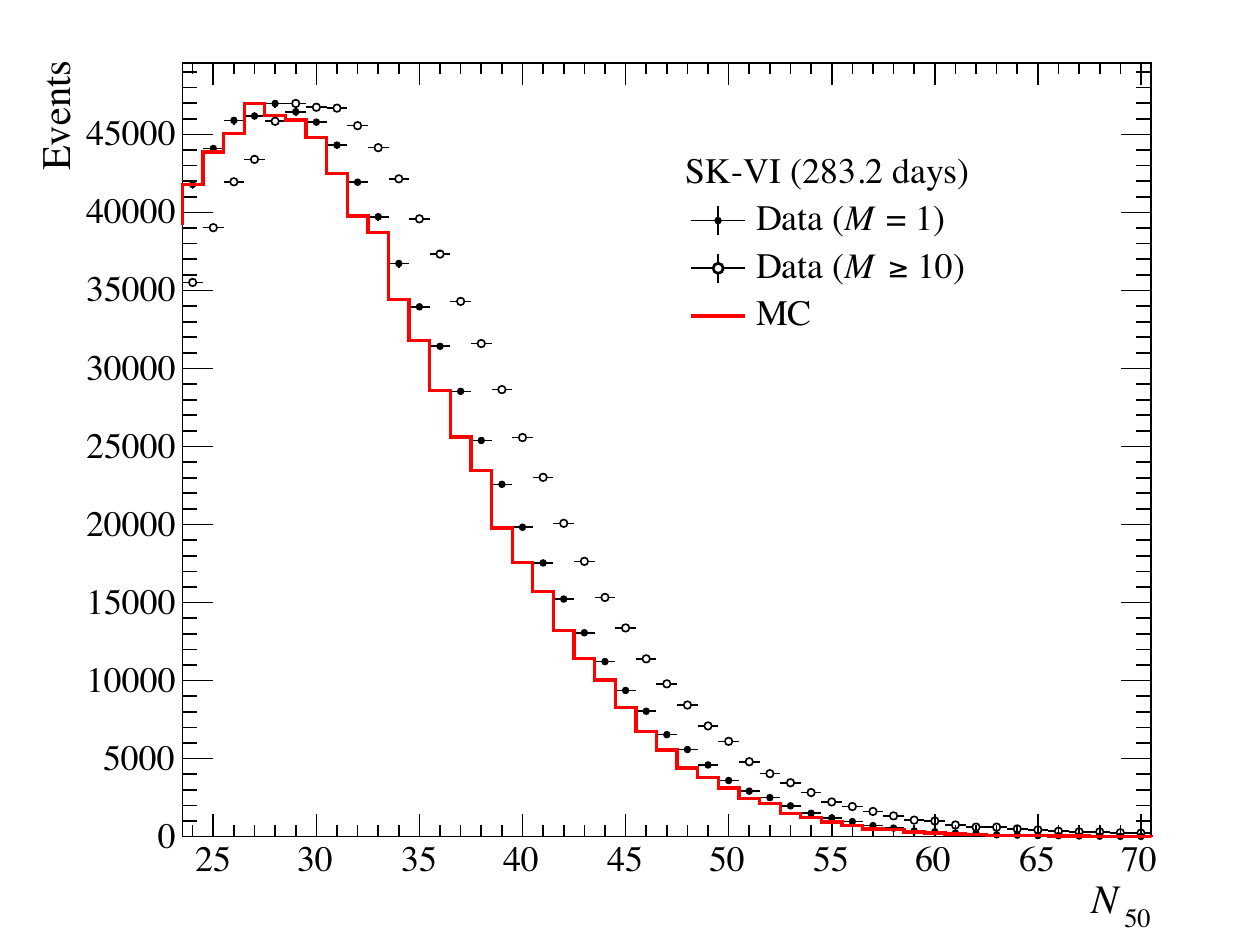}
    \caption{\label{fig:n50_data_1_ge10_mc} Comparison of $N_{50}$ distributions for MC (solid red line) with data for $M=1$ (filled circles) and $M \geq 10$ (open circles), where $M$ is the number of neutron capture candidates following a muon that remain after the event selections.}
\end{figure}

The largest source of the systematic uncertainty is the signal efficiency for the $N_{50}$ cut. In the $N_{50}$ distribution shown in Fig.~\ref{fig:n50}, there is a 4.1\% discrepancy in the $N_{50}$ scale between the data and the MC. This value is estimated by scaling the MC distribution to fit the data in the range $24 \leq N_{50} \leq 70$.
While neutrons are generated uniformly in the ID one at a time in the MC, several neutrons are produced at once from muon spallation and multiple neutron captures can occur simultaneously.
Because of this feature in the selection of neutron candidates within the same time window, the $N_{50}$ distribution for the data becomes larger than that for MC.
The correlation between the number of neutron capture events following a muon and the discrepancy in $N_{50}$ distributions between data and MC is shown in Fig.~\ref{fig:n50_data_1_ge10_mc}.
For candidates with single neutron capture the discrepancy is 1.6\%, while it is 7.9\% for events with greater than or equal to ten neutron captures. As the number of neutron candidates increases, the discrepancy increases due to the pile-up effect. This pile-up effect is also confirmed in the MC with multiple neutrons generated simultaneously.
As there is no reliable model to predict the number of neutron produced from the spallation of cosmic-ray muons at the underground detector, the systematic uncertainty of 4.7\% is assigned to account for the variation in the signal efficiency for the $N_{50}$ cut when the $N_{50}$ distribution is scaled by 4.1\%.

The fraction of neutron capture on hydrogen nuclei is 0.4\% in the range $24 \leq N_{50} \leq 70$ for the MC shown in Fig.~\ref{fig:n50}. This value is accounted for as the systematic uncertainty.

Leak-in/-out of neutrons is one of the main sources of the systematic uncertainty. The event rate of neutron capture around the center of the ID is expected to be uniform as the numbers of neutrons that leak in and out are likely balanced. On the other hand, the event rate near the ID wall is expected to be lower than that around the center. This is because when muons penetrate the surrounding rock or the OD, neutrons from their spallation are not counted even if they are reconstructed in the ID because the parent muons are not tagged in this analysis.
A systematic uncertainty of 2.2\% is assigned to account for the leak-in/-out by changing the boundary of the fiducial volume from 4 to 7\,m from the ID wall.

The uncertainty on the background estimation was estimated to be 0.3\%. This value is evaluated by fitting the $\Delta t$ distribution shown in Fig.~\ref{fig:dt_fit} while varying the parameter $B$ by $\pm 5.7$\%, which comes from the statistical uncertainty of the random trigger sample.

The total systematic uncertainty is estimated to be 6.7\% by adding all in quadrature.

\begin{table}
\caption{\label{tab:summary_syst}Summary of the systematic uncertainties.}
\begin{ruledtabular}
\begin{tabular}{lc}
    Source & Uncertainty (\%)\\
    \colrule
    Number of muons & 2.0\\
    Muon path length & 1.3\\
    Gd capture fraction & 2.2\\
    Gd capture time & 1.2\\
    Neutron thermalization & 2.2\\
    Signal efficiency for $L_t$ cut & 0.6\\
    Signal efficiency for $N_{50}$ cut & 4.7\\
    Contamination of hydrogen capture & 0.4\\
    Leak-in/-out of neutrons & 2.2\\
    Background estimation & 0.3\\
    \colrule
    Total & 6.7
\end{tabular}
\end{ruledtabular}
\end{table}

\section{\label{sec:results}Results}
The neutron yield $Y_n$ is defined as the neutron production rate per unit muon track length and per unit density, and can be calculated as
\begin{equation}
    Y_n = \frac{N_n}{N_\mu L_\mu \rho} = \frac{S_n}{\epsilon N_\mu L_\mu \rho},
\end{equation}
where $\rho$ is the density of the Gd sulfate solution, $\rho = 1.000\,\mathrm{g/cm^3}$. The uncertainty is negligible because it is smaller than 0.1\%. The other parameters are explained in the previous sections. It should be noted that the neutron yield includes both primary and secondary neutrons.
The neutron yield is measured to be $(2.76 \pm 0.02\,\mathrm{(stat.)} \pm 0.19\,\mathrm{(syst.)}) \times 10^{-4}\,\mu^{-1} \mathrm{g^{-1} cm^{2}}$.

\begin{figure}
    \centering
    \includegraphics[width=8.5cm]{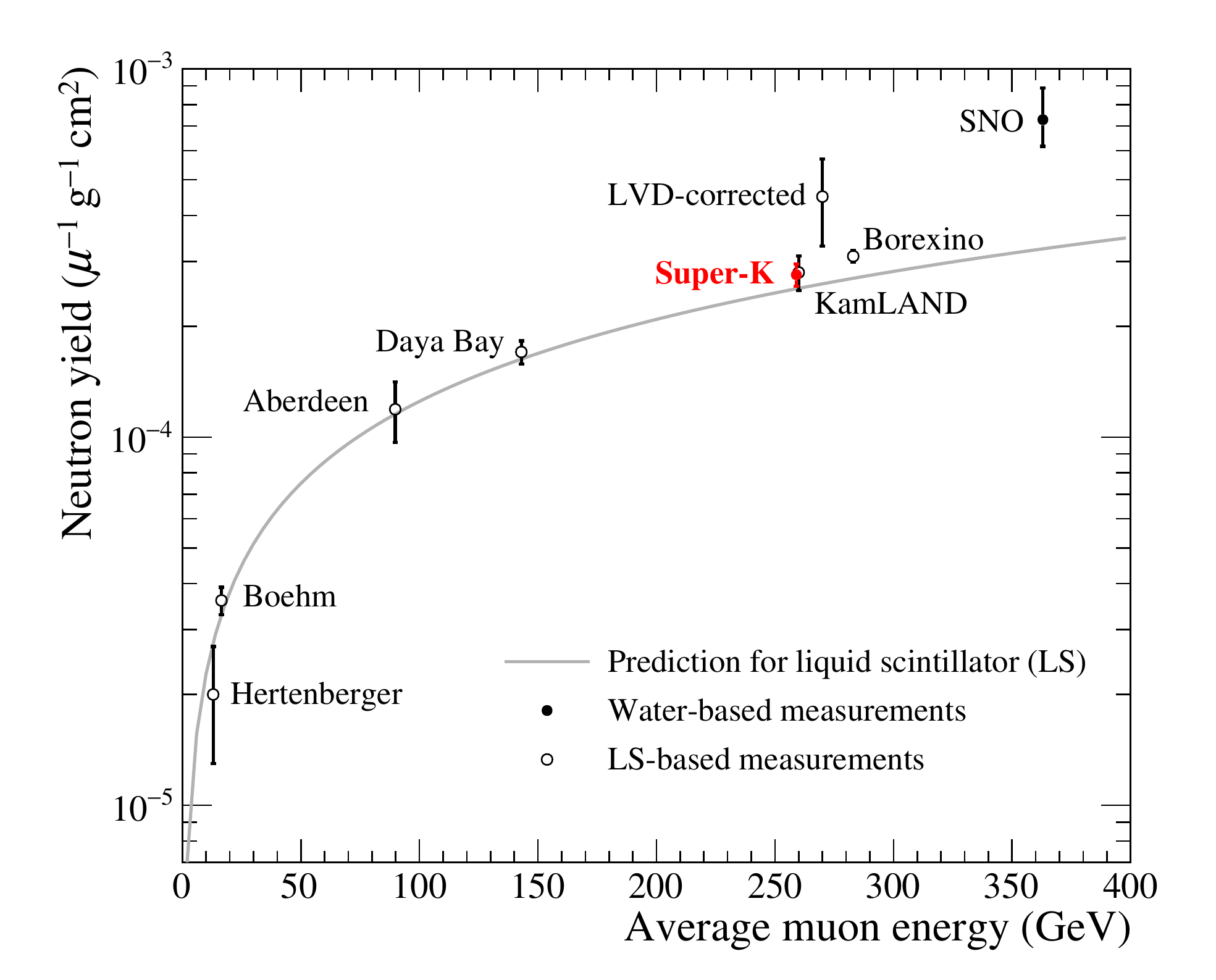}
    \caption{\label{fig:nyield_muene} Correlations between neutron yields and muon energies in various experiments. The muon energy corresponds to the depth at which the detector is located. The filled circles represent the water-based results \cite{sno}; in particular, the red one shows the result of this analysis. The open circles represent the measurements with liquid scintillator (LS) detector \cite{hertenberger,boehm,aberdeen,dayabay,kamland,lvd,borexino,mei_hime}. The gray solid line shows predictions for LS using FLUKA \cite{wang_etal}.}
\end{figure}

Comparisons with other experiments are shown in Fig.~\ref{fig:nyield_muene}.
Most of those yields were measured using liquid scintillators except for the SNO experiment which measured the yield in heavy water \cite{sno}. SK is the first experiment to measure the yield in light water.
The KamLAND detector is located in the same mountain as the SK detector at almost the same depth. Although the target material in the SK detector is water and different from the KamLAND detector, the measured neutron yields are consistent within the uncertainties.

In addition, neutron yields are calculated for each muon direction. Due to the shape of the mountains surrounding SK, the flux of muons from each azimuthal angle and the average energy at the detector site are different. The azimuthal dependence of the muon flux is shown in Fig.~\ref{fig:muon_flux}. The azimuthal distribution is divided into four parts of $90^\circ$ each, and the muon energy and neutron yield are estimated for each region. The results shown in Table~\ref{tab:azimutal_yield} can be interpreted as the higher the muon energy, the greater the neutron yield, as expected.

\begin{table}
\caption{\label{tab:azimutal_yield}Average muon energy and neutron yield for each azimuthal angle of muon travel. The units of the neutron yield are $10^{-4}\,\mu^{-1} \mathrm{g^{-1} cm^{2}}$. The error is statistical only.}
\begin{ruledtabular}
\begin{tabular}{cccc}
    Azimuthal angle & Fraction (\%) & $\overline{E}_\mu$\,(GeV) & $Y_n$ \\
    \colrule
    $315^\circ$--$45^\circ$ & 19.4 & 265 & $2.81 \pm 0.05$ \\
    $45^\circ$--$135^\circ$ & 11.5 & 272 & $2.88 \pm 0.06$ \\
    $135^\circ$--$225^\circ$ & 27.1 & 257 & $2.77 \pm 0.04$ \\
    $225^\circ$--$315^\circ$ & 42.0 & 253 & $2.67 \pm 0.03$
\end{tabular}
\end{ruledtabular}
\end{table}

\section{\label{sec:conclusion} Conclusion}
This paper presented the measurement of the neutron yield produced by cosmic-ray muon spallation in SK. The yield was found to be $(2.76 \pm 0.02\,\mathrm{(stat.)} \pm 0.19\,\mathrm{(syst.)}) \times 10^{-4}\,\mu^{-1} \mathrm{g^{-1} cm^{2}}$ at an average muon energy of 259\,GeV. The yield is consistent with the measurement by KamLAND at a similar depth, although the target material is different.  The energy dependence of the yield was also confirmed from a comparison for each muon direction.

As additional information, the Gd concentration was increased in July 2022 from 0.011 to 0.033\,wt\%.
This has increased the fraction of neutron capture by Gd to 75\%, and even more efficient measurements are expected for future analysis in SK.

\section{\label{sec:acknowledgement} Acknowledgement}
We gratefully acknowledge the cooperation of the Kamioka Mining and Smelting Company.
The Super‐Kamiokande experiment has been built and operated using funding by the 
Japanese Ministry of Education, Culture, Sports, Science and Technology, the U.S.
Department of Energy, and the U.S. National Science Foundation. Some of us have been 
supported by funds from the National Research Foundation of Korea NRF‐2009‐0083526
(KNRC) funded by the Ministry of Science, Information and Communication Technology (ICT), and Future Planning and the Ministry of
Education (No.~2018R1D1A1B07049158 and No.~2021R1I1A1A01059559), 
the Japan Society for the Promotion of Science, the National
Natural Science Foundation of China under Grant No.~11620101004, the Spanish Ministry of Science, 
Universities and Innovation (Grant No.~PGC2018-099388-B-I00), the Natural Sciences and 
Engineering Research Council (NSERC) of Canada, the Scinet and Westgrid consortia of
Compute Canada, the National Science Centre (No.~UMO-2018/30/E/ST2/00441) and the Ministry
of Education and Science (No.~DIR/WK/2017/05), Poland,
the Science and Technology Facilities Council (STFC) and Grid for Particle Physics (GridPP), UK, the European Union's 
Horizon 2020 Research and Innovation Programme under the Marie Sklodowska-Curie Grant
Agreement No.~754496, H2020-MSCA-RISE-2018 JENNIFER2 Grant Agreement No.~822070, and 
H2020-MSCA-RISE-2019 SK2HK Grant Agreement No.~872549.

\nocite{*}

\bibliography{apssamp}

\providecommand{\noopsort}[1]{}\providecommand{\singleletter}[1]{#1}%
\begin{thebibliography}{36}%
\makeatletter
\providecommand \@ifxundefined [1]{%
 \@ifx{#1\undefined}
}%
\providecommand \@ifnum [1]{%
 \ifnum #1\expandafter \@firstoftwo
 \else \expandafter \@secondoftwo
 \fi
}%
\providecommand \@ifx [1]{%
 \ifx #1\expandafter \@firstoftwo
 \else \expandafter \@secondoftwo
 \fi
}%
\providecommand \natexlab [1]{#1}%
\providecommand \enquote  [1]{``#1''}%
\providecommand \bibnamefont  [1]{#1}%
\providecommand \bibfnamefont [1]{#1}%
\providecommand \citenamefont [1]{#1}%
\providecommand \href@noop [0]{\@secondoftwo}%
\providecommand \href [0]{\begingroup \@sanitize@url \@href}%
\providecommand \@href[1]{\@@startlink{#1}\@@href}%
\providecommand \@@href[1]{\endgroup#1\@@endlink}%
\providecommand \@sanitize@url [0]{\catcode `\\12\catcode `\$12\catcode
  `\&12\catcode `\#12\catcode `\^12\catcode `\_12\catcode `\%12\relax}%
\providecommand \@@startlink[1]{}%
\providecommand \@@endlink[0]{}%
\providecommand \url  [0]{\begingroup\@sanitize@url \@url }%
\providecommand \@url [1]{\endgroup\@href {#1}{\urlprefix }}%
\providecommand \urlprefix  [0]{URL }%
\providecommand \Eprint [0]{\href }%
\providecommand \doibase [0]{https://doi.org/}%
\providecommand \selectlanguage [0]{\@gobble}%
\providecommand \bibinfo  [0]{\@secondoftwo}%
\providecommand \bibfield  [0]{\@secondoftwo}%
\providecommand \translation [1]{[#1]}%
\providecommand \BibitemOpen [0]{}%
\providecommand \bibitemStop [0]{}%
\providecommand \bibitemNoStop [0]{.\EOS\space}%
\providecommand \EOS [0]{\spacefactor3000\relax}%
\providecommand \BibitemShut  [1]{\csname bibitem#1\endcsname}%
\let\auto@bib@innerbib\@empty
\bibitem [{\citenamefont {Malgin}(2017)}]{malgin}%
  \BibitemOpen
  \bibfield  {author} {\bibinfo {author} {\bibfnamefont {A.~S.}\ \bibnamefont
  {Malgin}},\ }\href@noop {} {\bibfield  {journal} {\bibinfo  {journal} {Phys.\
  Rev.\ C}\ }\textbf {\bibinfo {volume} {96}},\ \bibinfo {pages} {014605}
  (\bibinfo {year} {2017})}\BibitemShut {NoStop}%
\bibitem [{\citenamefont {Luu}\ and\ \citenamefont
  {Hagmann}(2006)}]{Luu2006NeutronPB}%
  \BibitemOpen
  \bibfield  {author} {\bibinfo {author} {\bibfnamefont {T.}~\bibnamefont
  {Luu}}\ and\ \bibinfo {author} {\bibfnamefont {C.}~\bibnamefont {Hagmann}},\
  }\bibfield  {title} {\bibinfo {title} {Neutron {P}roduction by {M}uon
  {S}pallation {I}: Theory ({UCRLTR}- 226323)}\ }(\bibinfo {year}
  {2006})\BibitemShut {NoStop}%
\bibitem [{\citenamefont {Wang}\ \emph {et~al.}(2001)\citenamefont {Wang},
  \citenamefont {Balic}, \citenamefont {Gratta}, \citenamefont {Fass\`{o}},
  \citenamefont {Roesler},\ and\ \citenamefont {Ferrari}}]{wang_etal}%
  \BibitemOpen
  \bibfield  {author} {\bibinfo {author} {\bibfnamefont {Y.-F.}\ \bibnamefont
  {Wang}}, \bibinfo {author} {\bibfnamefont {V.}~\bibnamefont {Balic}},
  \bibinfo {author} {\bibfnamefont {G.}~\bibnamefont {Gratta}}, \bibinfo
  {author} {\bibfnamefont {A.}~\bibnamefont {Fass\`{o}}}, \bibinfo {author}
  {\bibfnamefont {S.}~\bibnamefont {Roesler}},\ and\ \bibinfo {author}
  {\bibfnamefont {A.}~\bibnamefont {Ferrari}},\ }\href@noop {} {\bibfield
  {journal} {\bibinfo  {journal} {Phys.\ Rev.\ D}\ }\textbf {\bibinfo {volume}
  {64}},\ \bibinfo {pages} {013012} (\bibinfo {year} {2001})}\BibitemShut
  {NoStop}%
\bibitem [{\citenamefont {Li}\ and\ \citenamefont
  {Beacom}(2014)}]{calc_li_beacom}%
  \BibitemOpen
  \bibfield  {author} {\bibinfo {author} {\bibfnamefont {S.~W.}\ \bibnamefont
  {Li}}\ and\ \bibinfo {author} {\bibfnamefont {J.~F.}\ \bibnamefont
  {Beacom}},\ }\href@noop {} {\bibfield  {journal} {\bibinfo  {journal} {Phys.\
  Rev.\ C}\ }\textbf {\bibinfo {volume} {89}},\ \bibinfo {pages} {045801}
  (\bibinfo {year} {2014})}\BibitemShut {NoStop}%
\bibitem [{\citenamefont {Li}\ and\ \citenamefont
  {Beacom}(2015)}]{calc_li_beacom2}%
  \BibitemOpen
  \bibfield  {author} {\bibinfo {author} {\bibfnamefont {S.~W.}\ \bibnamefont
  {Li}}\ and\ \bibinfo {author} {\bibfnamefont {J.~F.}\ \bibnamefont
  {Beacom}},\ }\href@noop {} {\bibfield  {journal} {\bibinfo  {journal} {Phys.\
  Rev.\ D}\ }\textbf {\bibinfo {volume} {91}},\ \bibinfo {pages} {105005}
  (\bibinfo {year} {2015})}\BibitemShut {NoStop}%
\bibitem [{\citenamefont {Hertenberger}\ \emph {et~al.}(1995)\citenamefont
  {Hertenberger}, \citenamefont {Chen},\ and\ \citenamefont
  {Dougherty}}]{hertenberger}%
  \BibitemOpen
  \bibfield  {author} {\bibinfo {author} {\bibfnamefont {R.}~\bibnamefont
  {Hertenberger}}, \bibinfo {author} {\bibfnamefont {M.}~\bibnamefont {Chen}},\
  and\ \bibinfo {author} {\bibfnamefont {B.~L.}\ \bibnamefont {Dougherty}},\
  }\href@noop {} {\bibfield  {journal} {\bibinfo  {journal} {Phys.\ Rev.\ C}\
  }\textbf {\bibinfo {volume} {52}},\ \bibinfo {pages} {3449} (\bibinfo {year}
  {1995})}\BibitemShut {NoStop}%
\bibitem [{\citenamefont {Boehm}\ \emph {et~al.}(2000)\citenamefont {Boehm}
  \emph {et~al.}}]{boehm}%
  \BibitemOpen
  \bibfield  {author} {\bibinfo {author} {\bibfnamefont {F.}~\bibnamefont
  {Boehm}} \emph {et~al.},\ }\href@noop {} {\bibfield  {journal} {\bibinfo
  {journal} {Phys.\ Rev.\ D}\ }\textbf {\bibinfo {volume} {62}},\ \bibinfo
  {pages} {092005} (\bibinfo {year} {2000})}\BibitemShut {NoStop}%
\bibitem [{\citenamefont {Blyth}\ \emph {et~al.}(2016)\citenamefont {Blyth}
  \emph {et~al.}}]{aberdeen}%
  \BibitemOpen
  \bibfield  {author} {\bibinfo {author} {\bibfnamefont {S.~C.}\ \bibnamefont
  {Blyth}} \emph {et~al.} (\bibinfo {collaboration} {Aberdeen Tunnel Experiment
  Collaboration}),\ }\href@noop {} {\bibfield  {journal} {\bibinfo  {journal}
  {Phys.\ Rev.\ D}\ }\textbf {\bibinfo {volume} {93}},\ \bibinfo {pages}
  {072005} (\bibinfo {year} {2016})}\BibitemShut {NoStop}%
\bibitem [{\citenamefont {An}\ \emph {et~al.}(2018)\citenamefont {An} \emph
  {et~al.}}]{dayabay}%
  \BibitemOpen
  \bibfield  {author} {\bibinfo {author} {\bibfnamefont {F.~P.}\ \bibnamefont
  {An}} \emph {et~al.} (\bibinfo {collaboration} {Daya Bay Collaboration}),\
  }\href@noop {} {\bibfield  {journal} {\bibinfo  {journal} {Phys.\ Rev.\ D}\
  }\textbf {\bibinfo {volume} {97}},\ \bibinfo {pages} {052009} (\bibinfo
  {year} {2018})}\BibitemShut {NoStop}%
\bibitem [{\citenamefont {Abe}\ \emph {et~al.}(2010)\citenamefont {Abe} \emph
  {et~al.}}]{kamland}%
  \BibitemOpen
  \bibfield  {author} {\bibinfo {author} {\bibfnamefont {S.}~\bibnamefont
  {Abe}} \emph {et~al.} (\bibinfo {collaboration} {KamLAND Collaboration}),\
  }\href@noop {} {\bibfield  {journal} {\bibinfo  {journal} {Phys.\ Rev.\ C}\
  }\textbf {\bibinfo {volume} {81}},\ \bibinfo {pages} {025807} (\bibinfo
  {year} {2010})}\BibitemShut {NoStop}%
\bibitem [{\citenamefont {Aglietta}\ \emph {et~al.}(1999)\citenamefont
  {Aglietta} \emph {et~al.}}]{lvd}%
  \BibitemOpen
  \bibinfo {editor} {\bibfnamefont {M.}~\bibnamefont {Aglietta}} \emph
  {et~al.},\ eds.,\ \href@noop {} {\emph {\bibinfo {title} {26th International
  Cosmic Ray Conference}}}\ (\bibinfo  {publisher} {AIP Conference
  Proceedings},\ \bibinfo {year} {1999})\BibitemShut {NoStop}%
\bibitem [{\citenamefont {Bellini}\ \emph {et~al.}(2013)\citenamefont {Bellini}
  \emph {et~al.}}]{borexino}%
  \BibitemOpen
  \bibfield  {author} {\bibinfo {author} {\bibfnamefont {G.}~\bibnamefont
  {Bellini}} \emph {et~al.} (\bibinfo {collaboration} {Borexino
  Collaboration}),\ }\href@noop {} {\bibfield  {journal} {\bibinfo  {journal}
  {J.\ Cosmol.\ Astropart.\ Phys. 08}\ ,\ \bibinfo {pages} {049}} (\bibinfo
  {year} {2013})}\BibitemShut {NoStop}%
\bibitem [{\citenamefont {Fukuda}\ \emph {et~al.}(2003)\citenamefont {Fukuda}
  \emph {et~al.}}]{Super-Kamiokande:2002weg}%
  \BibitemOpen
  \bibfield  {author} {\bibinfo {author} {\bibfnamefont {Y.}~\bibnamefont
  {Fukuda}} \emph {et~al.} (\bibinfo {collaboration} {Super-Kamiokande
  Collaboration}),\ }\href@noop {} {\bibfield  {journal} {\bibinfo  {journal}
  {Nucl. Instrum. Meth. A}\ }\textbf {\bibinfo {volume} {501}},\ \bibinfo
  {pages} {418} (\bibinfo {year} {2003})}\BibitemShut {NoStop}%
\bibitem [{\citenamefont {Abe}\ \emph {et~al.}(2022{\natexlab{a}})\citenamefont
  {Abe} \emph {et~al.}}]{sk_gd_loading}%
  \BibitemOpen
  \bibfield  {author} {\bibinfo {author} {\bibfnamefont {K.}~\bibnamefont
  {Abe}} \emph {et~al.} (\bibinfo {collaboration} {Super-Kamiokande
  Collaboration}),\ }\href@noop {} {\bibfield  {journal} {\bibinfo  {journal}
  {Nucl. Instrum. Methods Phys. Res., Sect. A}\ }\textbf {\bibinfo {volume}
  {1027}},\ \bibinfo {pages} {166248} (\bibinfo {year}
  {2022}{\natexlab{a}})}\BibitemShut {NoStop}%
\bibitem [{\citenamefont {Yamada}\ \emph {et~al.}(2010)\citenamefont {Yamada}
  \emph {et~al.}}]{sk_qbee}%
  \BibitemOpen
  \bibfield  {author} {\bibinfo {author} {\bibfnamefont {S.}~\bibnamefont
  {Yamada}} \emph {et~al.},\ }\href@noop {} {\bibfield  {journal} {\bibinfo
  {journal} {IEEE Transactions on Nuclear Science}\ }\textbf {\bibinfo {volume}
  {57}},\ \bibinfo {pages} {428} (\bibinfo {year} {2010})}\BibitemShut
  {NoStop}%
\bibitem [{\citenamefont {Zhang}\ \emph {et~al.}(2016)\citenamefont {Zhang}
  \emph {et~al.}}]{sk_li9}%
  \BibitemOpen
  \bibfield  {author} {\bibinfo {author} {\bibfnamefont {Y.}~\bibnamefont
  {Zhang}} \emph {et~al.} (\bibinfo {collaboration} {Super-Kamiokande
  Collaboration}),\ }\href@noop {} {\bibfield  {journal} {\bibinfo  {journal}
  {Phys.\ Rev.\ D}\ }\textbf {\bibinfo {volume} {93}},\ \bibinfo {pages}
  {012004} (\bibinfo {year} {2016})}\BibitemShut {NoStop}%
\bibitem [{\citenamefont {Watanabe}\ \emph {et~al.}(2009)\citenamefont
  {Watanabe} \emph {et~al.}}]{sk_1st_ntag}%
  \BibitemOpen
  \bibfield  {author} {\bibinfo {author} {\bibfnamefont {H.}~\bibnamefont
  {Watanabe}} \emph {et~al.} (\bibinfo {collaboration} {Super-Kamiokande
  Collaboration}),\ }\href@noop {} {\bibfield  {journal} {\bibinfo  {journal}
  {Astropart. Phys.}\ }\textbf {\bibinfo {volume} {31}},\ \bibinfo {pages}
  {320} (\bibinfo {year} {2009})}\BibitemShut {NoStop}%
\bibitem [{\citenamefont {Abe}\ \emph {et~al.}(2021)\citenamefont {Abe} \emph
  {et~al.}}]{sk4_srn}%
  \BibitemOpen
  \bibfield  {author} {\bibinfo {author} {\bibfnamefont {K.}~\bibnamefont
  {Abe}} \emph {et~al.} (\bibinfo {collaboration} {Super-Kamiokande
  Collaboration}),\ }\href@noop {} {\bibfield  {journal} {\bibinfo  {journal}
  {Phys.\ Rev.\ D}\ }\textbf {\bibinfo {volume} {104}},\ \bibinfo {pages}
  {122002} (\bibinfo {year} {2021})}\BibitemShut {NoStop}%
\bibitem [{\citenamefont {Abe}\ \emph {et~al.}(2022{\natexlab{b}})\citenamefont
  {Abe} \emph {et~al.}}]{ntag_sk_2022}%
  \BibitemOpen
  \bibfield  {author} {\bibinfo {author} {\bibfnamefont {K.}~\bibnamefont
  {Abe}} \emph {et~al.} (\bibinfo {collaboration} {Super-Kamiokande
  Collaboration}),\ }\href@noop {} {\bibfield  {journal} {\bibinfo  {journal}
  {JINST\,}\ }\textbf {\bibinfo {volume} {17}},\ \bibinfo {pages} {P10029}
  (\bibinfo {year} {2022}{\natexlab{b}})}\BibitemShut {NoStop}%
\bibitem [{\citenamefont {Brun}\ \emph {et~al.}(1994)\citenamefont {Brun} \emph
  {et~al.}}]{GEANT3}%
  \BibitemOpen
  \bibfield  {author} {\bibinfo {author} {\bibfnamefont {R.}~\bibnamefont
  {Brun}} \emph {et~al.},\ }\href@noop {} {\bibfield  {journal} {\bibinfo
  {journal} {Report No. CERN-W5013}\ } (\bibinfo {year} {1994})}\BibitemShut
  {NoStop}%
\bibitem [{\citenamefont {Abe}\ \emph {et~al.}(2014)\citenamefont {Abe} \emph
  {et~al.}}]{sk_calib}%
  \BibitemOpen
  \bibfield  {author} {\bibinfo {author} {\bibfnamefont {K.}~\bibnamefont
  {Abe}} \emph {et~al.} (\bibinfo {collaboration} {Super-Kamiokande
  Collaboration}),\ }\href@noop {} {\bibfield  {journal} {\bibinfo  {journal}
  {Nucl. Instrum. Methods Phys. Res., Sect. A}\ }\textbf {\bibinfo {volume}
  {737}},\ \bibinfo {pages} {253} (\bibinfo {year} {2014})}\BibitemShut
  {NoStop}%
\bibitem [{\citenamefont {Agostinelli}\ \emph {et~al.}(2003)\citenamefont
  {Agostinelli} \emph {et~al.}}]{GEANT4:2002zbu}%
  \BibitemOpen
  \bibfield  {author} {\bibinfo {author} {\bibfnamefont {S.}~\bibnamefont
  {Agostinelli}} \emph {et~al.},\ }\href@noop {} {\bibfield  {journal}
  {\bibinfo  {journal} {Nucl. Instrum. Meth. A}\ }\textbf {\bibinfo {volume}
  {506}},\ \bibinfo {pages} {250} (\bibinfo {year} {2003})}\BibitemShut
  {NoStop}%
\bibitem [{\citenamefont {Ou}\ \emph {et~al.}(2014)\citenamefont {Ou} \emph
  {et~al.}}]{annri_gd}%
  \BibitemOpen
  \bibfield  {author} {\bibinfo {author} {\bibfnamefont {I.}~\bibnamefont {Ou}}
  \emph {et~al.},\ }\href@noop {} {\bibfield  {journal} {\bibinfo  {journal}
  {AIP Conf. Proc.}\ }\textbf {\bibinfo {volume} {1594}},\ \bibinfo {pages}
  {351} (\bibinfo {year} {2014})}\BibitemShut {NoStop}%
\bibitem [{\citenamefont {Conner}(2004)}]{muboy_conner}%
  \BibitemOpen
  \bibfield  {author} {\bibinfo {author} {\bibfnamefont {Z.}~\bibnamefont
  {Conner}},\ }\emph {\bibinfo {title} {A Study of solar neutrinos using the
  Super-Kamiokande Detector}},\ \href@noop {} {Ph.D. thesis},\ \bibinfo
  {school} {Boston University} (\bibinfo {year} {2004})\BibitemShut {NoStop}%
\bibitem [{\citenamefont {Desai}(2004)}]{muboy_desai}%
  \BibitemOpen
  \bibfield  {author} {\bibinfo {author} {\bibfnamefont {S.}~\bibnamefont
  {Desai}},\ }\emph {\bibinfo {title} {High energy neutrino astrophysics with
  Super-Kamiokande}},\ \href@noop {} {Ph.D. thesis},\ \bibinfo  {school}
  {Boston University} (\bibinfo {year} {2004})\BibitemShut {NoStop}%
\bibitem [{\citenamefont {Gaisser}\ and\ \citenamefont
  {Stanev}(2004)}]{gaisser}%
  \BibitemOpen
  \bibfield  {author} {\bibinfo {author} {\bibfnamefont {T.~K.}\ \bibnamefont
  {Gaisser}}\ and\ \bibinfo {author} {\bibfnamefont {T.}~\bibnamefont
  {Stanev}},\ }\href@noop {} {\bibfield  {journal} {\bibinfo  {journal} {Phys.
  Lett. B}\ }\textbf {\bibinfo {volume} {592}},\ \bibinfo {pages} {228}
  (\bibinfo {year} {2004})}\BibitemShut {NoStop}%
\bibitem [{\citenamefont {Tang}\ \emph {et~al.}(2006)\citenamefont {Tang},
  \citenamefont {Horton-Smith}, \citenamefont {Kudryavtsev},\ and\
  \citenamefont {Tonazzo}}]{tang}%
  \BibitemOpen
  \bibfield  {author} {\bibinfo {author} {\bibfnamefont {A.}~\bibnamefont
  {Tang}}, \bibinfo {author} {\bibfnamefont {G.}~\bibnamefont {Horton-Smith}},
  \bibinfo {author} {\bibfnamefont {V.~A.}\ \bibnamefont {Kudryavtsev}},\ and\
  \bibinfo {author} {\bibfnamefont {A.}~\bibnamefont {Tonazzo}},\ }\href@noop
  {} {\bibfield  {journal} {\bibinfo  {journal} {Phys.\ Rev.\ D}\ }\textbf
  {\bibinfo {volume} {74}},\ \bibinfo {pages} {053007} (\bibinfo {year}
  {2006})}\BibitemShut {NoStop}%
\bibitem [{\citenamefont {Antonioli}\ \emph {et~al.}(1997)\citenamefont
  {Antonioli}, \citenamefont {Ghetti}, \citenamefont {Korolkova}, \citenamefont
  {Kudryavtsev},\ and\ \citenamefont {Sartorelli}}]{music}%
  \BibitemOpen
  \bibfield  {author} {\bibinfo {author} {\bibfnamefont {P.}~\bibnamefont
  {Antonioli}}, \bibinfo {author} {\bibfnamefont {C.}~\bibnamefont {Ghetti}},
  \bibinfo {author} {\bibfnamefont {E.~V.}\ \bibnamefont {Korolkova}}, \bibinfo
  {author} {\bibfnamefont {V.~A.}\ \bibnamefont {Kudryavtsev}},\ and\ \bibinfo
  {author} {\bibfnamefont {G.}~\bibnamefont {Sartorelli}},\ }\href@noop {}
  {\bibfield  {journal} {\bibinfo  {journal} {Astropart.\ Phys.}\ }\textbf
  {\bibinfo {volume} {7}},\ \bibinfo {pages} {357} (\bibinfo {year}
  {1997})}\BibitemShut {NoStop}%
\bibitem [{dig(1997)}]{diggitalmap}%
  \BibitemOpen
  \bibfield  {title} {\bibinfo {title} {{G}eographical {S}urvey {I}nstitute of
  {J}apan, unpublished}} (\bibinfo {year} {1997})\BibitemShut {NoStop}%
\bibitem [{\citenamefont {Groom}\ \emph {et~al.}(2001)\citenamefont {Groom},
  \citenamefont {Mokhov},\ and\ \citenamefont {Striganov}}]{groom}%
  \BibitemOpen
  \bibfield  {author} {\bibinfo {author} {\bibfnamefont {D.~E.}\ \bibnamefont
  {Groom}}, \bibinfo {author} {\bibfnamefont {N.~V.}\ \bibnamefont {Mokhov}},\
  and\ \bibinfo {author} {\bibfnamefont {S.~I.}\ \bibnamefont {Striganov}},\
  }\href@noop {} {\bibfield  {journal} {\bibinfo  {journal} {Atomic Data and
  Nuclear Data Tables}\ }\textbf {\bibinfo {volume} {78}},\ \bibinfo {pages}
  {183} (\bibinfo {year} {2001})}\BibitemShut {NoStop}%
\bibitem [{\citenamefont {Barrett}\ \emph {et~al.}(1952)\citenamefont
  {Barrett}, \citenamefont {Bollinger}, \citenamefont {Cocconi}, \citenamefont
  {Eisenberg},\ and\ \citenamefont {Greisen}}]{barrett}%
  \BibitemOpen
  \bibfield  {author} {\bibinfo {author} {\bibfnamefont {P.~H.}\ \bibnamefont
  {Barrett}}, \bibinfo {author} {\bibfnamefont {L.~M.}\ \bibnamefont
  {Bollinger}}, \bibinfo {author} {\bibfnamefont {G.}~\bibnamefont {Cocconi}},
  \bibinfo {author} {\bibfnamefont {Y.}~\bibnamefont {Eisenberg}},\ and\
  \bibinfo {author} {\bibfnamefont {K.}~\bibnamefont {Greisen}},\ }\href@noop
  {} {\bibfield  {journal} {\bibinfo  {journal} {Rev. Mod. Phys.}\ }\textbf
  {\bibinfo {volume} {24}},\ \bibinfo {pages} {133} (\bibinfo {year}
  {1952})}\BibitemShut {NoStop}%
\bibitem [{\citenamefont {Nakano}\ \emph {et~al.}(2020)\citenamefont {Nakano}
  \emph {et~al.}}]{sk_nakano_Rn}%
  \BibitemOpen
  \bibfield  {author} {\bibinfo {author} {\bibfnamefont {Y.}~\bibnamefont
  {Nakano}} \emph {et~al.},\ }\href@noop {} {\bibfield  {journal} {\bibinfo
  {journal} {Nucl. Instrum. Methods Phys. Res., Sect. A}\ }\textbf {\bibinfo
  {volume} {977}},\ \bibinfo {pages} {164297} (\bibinfo {year}
  {2020})}\BibitemShut {NoStop}%
\bibitem [{\citenamefont {Smy}(2008)}]{smy_proc}%
  \BibitemOpen
  \bibfield  {author} {\bibinfo {author} {\bibfnamefont {M.}~\bibnamefont
  {Smy}},\ }\href@noop {} {\bibfield  {journal} {\bibinfo  {journal}
  {Proceeding of the 30th International Cosmic Ray Conference}\ }\textbf
  {\bibinfo {volume} {5}},\ \bibinfo {pages} {1279} (\bibinfo {year}
  {2008})}\BibitemShut {NoStop}%
\bibitem [{\citenamefont {Koshio}(1998)}]{koshio}%
  \BibitemOpen
  \bibfield  {author} {\bibinfo {author} {\bibfnamefont {Y.}~\bibnamefont
  {Koshio}},\ }\emph {\bibinfo {title} {Study of Solar Neutrinos at
  Super-Kamiokande}},\ \href@noop {} {Ph.D. thesis},\ \bibinfo  {school}
  {University of Tokyo} (\bibinfo {year} {1998})\BibitemShut {NoStop}%
\bibitem [{\citenamefont {Aharmim}\ \emph {et~al.}(2019)\citenamefont {Aharmim}
  \emph {et~al.}}]{sno}%
  \BibitemOpen
  \bibfield  {author} {\bibinfo {author} {\bibfnamefont {B.}~\bibnamefont
  {Aharmim}} \emph {et~al.} (\bibinfo {collaboration} {SNO Collaboration}),\
  }\href@noop {} {\bibfield  {journal} {\bibinfo  {journal} {Phys.\ Rev.\ D}\
  }\textbf {\bibinfo {volume} {100}},\ \bibinfo {pages} {112005} (\bibinfo
  {year} {2019})}\BibitemShut {NoStop}%
\bibitem [{\citenamefont {Mei}\ and\ \citenamefont {Hime}(2006)}]{mei_hime}%
  \BibitemOpen
  \bibfield  {author} {\bibinfo {author} {\bibfnamefont {D.-M.}\ \bibnamefont
  {Mei}}\ and\ \bibinfo {author} {\bibfnamefont {A.}~\bibnamefont {Hime}},\
  }\href@noop {} {\bibfield  {journal} {\bibinfo  {journal} {Phys.\ Rev.\ D}\
  }\textbf {\bibinfo {volume} {73}},\ \bibinfo {pages} {053004} (\bibinfo
  {year} {2006})}\BibitemShut {NoStop}%
\end{thebibliography}%

\end{document}